\documentclass[final,3p]{elsarticle} 

\usepackage{amssymb,amsfonts,amsmath,bm}
\usepackage{graphicx}
\usepackage{booktabs}
\usepackage[colorlinks=true,linkcolor=blue, citecolor=blue, urlcolor=blue,letterpaper,bookmarks=true,bookmarksopen=false,bookmarksnumbered=false,hypertexnames=true]{hyperref}

\biboptions{sort&compress}

\newcommand{\pd}[1]{\partial_{#1}}
\newcommand{\bL}{\text{\bf L}}
\newcommand{\bM}{\text{\bf M}}
\newcommand{\ov}[1]{\overline{#1}}
\newcommand{\bu}{\bm{u}}
\newcommand{\bbu}{\ov{\bu}}
\newcommand{\bbU}{\ov{\bm{U}}}
\newcommand{\oq}{\ov{q}}
\newcommand{\opsi}{\ov{\psi}}
\newcommand{\tnabla}{\tilde{\nabla}}
\newcommand{\eps}{\epsilon}

\journal{J. Comp. Phys.}

\begin{document}

\begin{frontmatter}
\title{Stochastic Superparameterization in Quasigeostrophic Turbulence}
\author[NYU]{Ian Grooms\corref{cor1}}\ead{grooms@cims.nyu.edu}\cortext[cor1]{Corresponding author}
\author[NYU,NYUAD]{Andrew J. Majda}\ead{jonjon@cims.nyu.edu}
\address[NYU]{Center for Atmosphere Ocean Science,
Courant Institute of Mathematical Sciences,
	 New York University, 251 Mercer St., New York, NY 10012. }
\address[NYUAD]{Center for Prototype Climate Modelling, 
NYU-Abu Dhabi}
\date{\today}

\begin{abstract}
In this article we expand and develop the authors' recent proposed methodology for efficient stochastic superparameterization algorithms for geophysical turbulence.
Geophysical turbulence is characterized by significant intermittent cascades of energy from the unresolved to the resolved scales resulting in complex patterns of waves, jets, and vortices.
Conventional superparameterization simulates large scale dynamics on a coarse grid in a physical domain, and couples these dynamics to high-resolution simulations on periodic domains embedded in the coarse grid.
Stochastic superparameterization replaces the nonlinear, deterministic eddy equations on periodic embedded domains by quasilinear stochastic approximations on formally infinite embedded domains.
The result is a seamless algorithm which never uses a small scale grid and is far cheaper than conventional SP, but with significant success in difficult test problems.

Various design choices in the algorithm are investigated in detail here, including decoupling the timescale of evolution on the embedded domains from the length of the time step used on the coarse grid, and sensitivity to certain assumed properties of the eddies (e.g.\,the shape of the assumed eddy energy spectrum).
We present four closures based on stochastic superparameterization which elucidate the properties of the underlying framework: a `null hypothesis' stochastic closure that uncouples the eddies from the mean, a stochastic closure with nonlinearly coupled eddies and mean, a nonlinear deterministic closure, and a stochastic closure based on energy conservation.
The different algorithms are compared and contrasted on a stringent test suite for quasigeostrophic turbulence involving two-layer dynamics on a $\beta$-plane forced by an imposed background shear.

The success of the algorithms developed here suggests that they may be fruitfully applied to more realistic situations.
They are expected to be particularly useful in providing accurate and efficient stochastic parameterizations for use in ensemble-based state estimation and prediction.
\end{abstract}


\end{frontmatter}

\section{Introduction}
\label{sec:Intro}
Computational physics often faces the challenge of simulating phenomena with complex interactions across a range of scales too wide to be accessible with existing supercomputers.
This is the case, for example, in simulations of global-scale atmospheric and oceanic dynamics, of solar magnetohydrodynamics, and of mantle convection, to name a few.
In these situations it is of paramount importance to provide accurate and efficient parameterizations of the effects of unresolved scales.
A novel approach combining elements of superparameterization \citep{GS99,KRD05,Taoetal09,CHJM11} and stochastic parameterization has been recently proposed by \citet{GM13a} (hereafter GM) and \citep{GM14}.

Superparameterization (SP) is a multiscale algorithm that was originally developed for the purpose of parameterizing unresolved cloud process in tropical atmospheric convection \citep{GS99,G01}.
In SP, high resolution simulations are embedded within the grid cells of a low resolution, large-scale model, to which they are coupled.
In the atmospheric context, the high resolution embedded domains share the vertical coordinate with the low resolution physical domain while the horizontal coordinates of the embedded domains are periodic so that the embedded domains in different coarse-grid cells are not directly coupled (some alternatives are discussed by \cite{Arakawa04,JA10}).
To reduce the computational expense of running an array of high resolution simulations, the embedded domains are usually made two-dimensional, with one horizontal and one vertical coordinate.
The computational expense can be further reduced by making the embedded domains smaller than the spatiotemporal grid of the coarse model \citep{XMG09}, or by embedding domains in a reduced number of coarse-grid cells \citep{MCJ13}.
Despite these innovations the computational cost of SP remains high compared to most alternative parameterizations.

\citet{MG09} proposed test models for SP, not as models of any particular process, but as a simplified setting for analyzing the mathematical structure of SP algorithms.
In these test models the small-scale dynamics on the embedded domains obey quasilinear stochastic partial differential equations with coefficients that are functions of the local large-scale variables. 
It is possible to compute the eddy feedback to the large scales very efficiently in these models without the need to directly simulate the small-scale dynamics on periodic embedded domains.
This idea of quasilinear stochastic eddies, coupled with the success of the above-mentioned SP algorithms that radically simplify the small-scale dynamics by making them two-dimensional, suggests that SP might be made still more efficient by replacing the nonlinear dynamics on the embedded domains with a quasilinear stochastic approximation.

This avenue was first proposed and implemented in a one-dimensional turbulent test problem in \citep{GM14}, where the algorithm met with resounding success in a complex situation with coherent solitons, dispersive waves, and an inverse cascade of energy from the unresolved small scales, punctuated by strong intermittent bursts of downscale cascade and dissipation associated with collapsing unstable solitons.
Although the small-scale dynamics in \citep{GM14} are stochastic, only the mean value of the eddy feedback to the large scales is used, so the eddy feedback terms are in fact non-stochastic.
In GM the authors further developed stochastic SP in a paradigm model of geophysical turbulence, namely two-layer quasigeostrophic (QG) turbulence on a $\beta$-plane forced by imposed zonal baroclinic shear.
GM expands the foundation of the SP algorithm in \citep{GM14} by developing a closure based on randomly oriented plane waves, making the feedback to the large scale stochastic, in contrast to \citep{GM14}; the initial results presented in GM show significant promise with skill in reproducing the inverse cascade of QG turbulence, and the jets and heat flux on the large scales in a variety of different regimes.
This paper develops the framework of GM in greater detail, and explores the parameter dependence of the method.
We also provide results for three new closures not presented in GM: a `null hypothesis' in which the eddy dynamics are decoupled from the large scales, a nonlinear deterministic closure not based on random plane waves, and a closure that predicts the key tunable parameter from GM based on energy conservation ideas.

The outline of the paper is as follows.
In section \ref{sec:DNS} the test problem is described in more detail, including the results of high-resolution reference simulations in three different parameter regimes.
In section \ref{sec:PA} we develop a multiscale framework, the `point approximation,' that allows us to move from a single set of equations governing the dynamics at all scales, to a set of coupled equations describing the large-scale dynamics on the physical domain and the small-scale dynamics on the embedded domains.
The stochastic approximation of the eddy equations is developed in section \ref{sec:GC}, and the four closures are presented in sections \ref{sec:ThreeClosures} and \ref{sec:EnergyMethods}.
Results of the numerical experiments are presented in section \ref{sec:Results}, and discussed in section \ref{sec:Discussion}.
Section \ref{sec:Conclusions} concludes.

\section{Reference Simulations\label{sec:DNS}}
\begin{figure}
\begin{center}
\includegraphics[width=\textwidth]{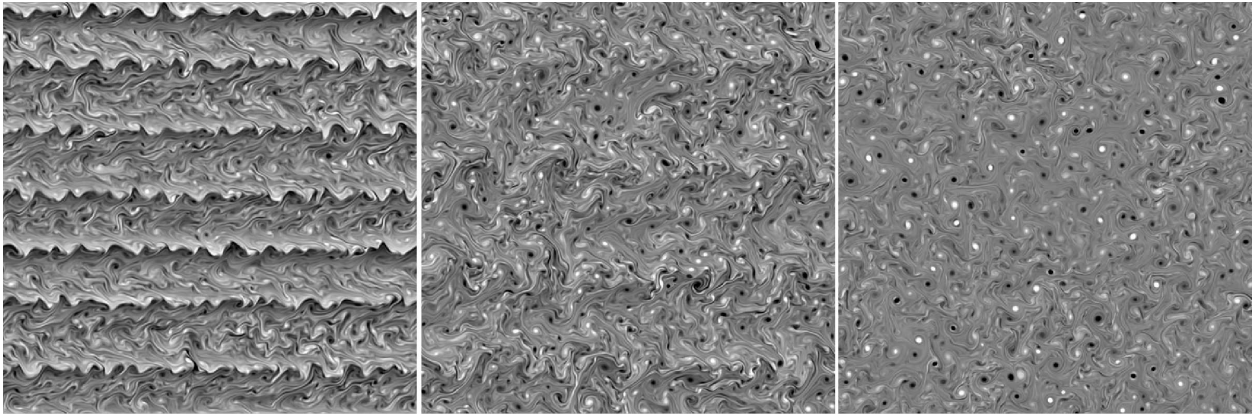}
\caption{Snapshots of barotropic potential vorticity $q_t$ from the high resolution reference simulations at weak (left), moderate (center), and strong supercriticalities (right).}\label{fig:DNS}
\end{center}
\end{figure}
As in GM, we test stochastic SP for two-equal-layer, rigid-lid, quasigeostrophic turbulence forced by an imposed, baroclinically unstable, horizontally uniform, vertically sheared zonal (east-west; $x$-direction) flow.
In this section we summarize the relevant properties of high-resolution direct numerical simulations (DNS) to serve as a reference.

The governing equations are
\begin{gather}\notag
\pd{t}q_1=-\nabla\cdot(\bu_1q_1)-\pd{x}q_1-(k_\beta^2 + k_d^2)v_1-\nu\nabla^8q_1,\\\notag
\pd{t}q_2=-\nabla\cdot(\bu_2q_2)+\pd{x}q_2-(k_\beta^2 - k_d^2)v_2-r\nabla^2\psi_2-\nu\nabla^8q_2\\\notag
q_1 = \nabla^2\psi_1+\frac{k_d^2}{2}(\psi_2-\psi_1),\\
q_2 = \nabla^2\psi_2-\frac{k_d^2}{2}(\psi_2-\psi_1),\label{eqn:QGPV}
\end{gather}
where $q_j$ is the potential vorticity in the upper ($j=1$) and lower ($j=2$) layers, $\nabla^2\psi_j=\omega_j$ is the relative vorticity, the velocity-streamfunction relation is $u_j=-\pd{y}\psi_j$, $v_j = \pd{x}\psi_j$, $k_d$ is the deformation wavenumber ($k_d^{-1}$ is the deformation radius), the coefficient $r$ specifies the strength of linear bottom friction (Ekman drag) and $\nu$ is the hyperviscous Reynolds number.
The terms $(k_\beta^2 + k_d^2)v_1$ in the upper layer and $(k_\beta^2 - k_d^2)v_2$ in the lower layer represent advection acting against an imposed large-scale meridional (north-south, $y$-direction) potential vorticity gradient; the $k_\beta^2$ terms result from the variation of the vertical projection of Coriolis frequency with latitude, and the $k_d^2$ terms result from the imposed vertical shear.
The dynamics can also be described in terms of barotropic and baroclinic modes, the former being given by the vertical average $q_t = (q_1+q_2)/2 = \nabla^2\psi_t$ and the latter by the vertical difference $q_c = (q_1-q_2)/2 = (\nabla^2-k_d^2)\psi_c$.
Subscripts $t$ and $c$ are used throughout to denote barotropic and baroclinic components, respectively.

Our reference solutions use $k_d=50$ and $512$ points in each direction of a square periodic domain of nondimensional width $2\pi$, which equals the highest resolution used in \citet{TY06,TY07} and minimally resolves the deformation scale with ten points per deformation wavelength.
In all three simulations the hyperviscous coefficient is $\nu = 1.5\times10^{-16}$; the nonlinear advection terms are dealiased using the $3/2$-rule, which means that they are equivalent to simulations at $768^2$ using the $2/3$-rule \citep{Orszag71}.
Time integration is via the adaptive, fourth-order, semi-implicit Runge-Kutta time integration scheme ARK4(3)6[L]2SA of \citet{KC03}, treating the hyperviscous terms implicitly, with PI.3.4 adaptive stepsize control based on error-per-step in the infinity norm on $q_j$ with a tolerance of $0.1$ \citep{Soderlind02}.

\begin{figure}[t]
\begin{center}
\includegraphics[width=\textwidth]{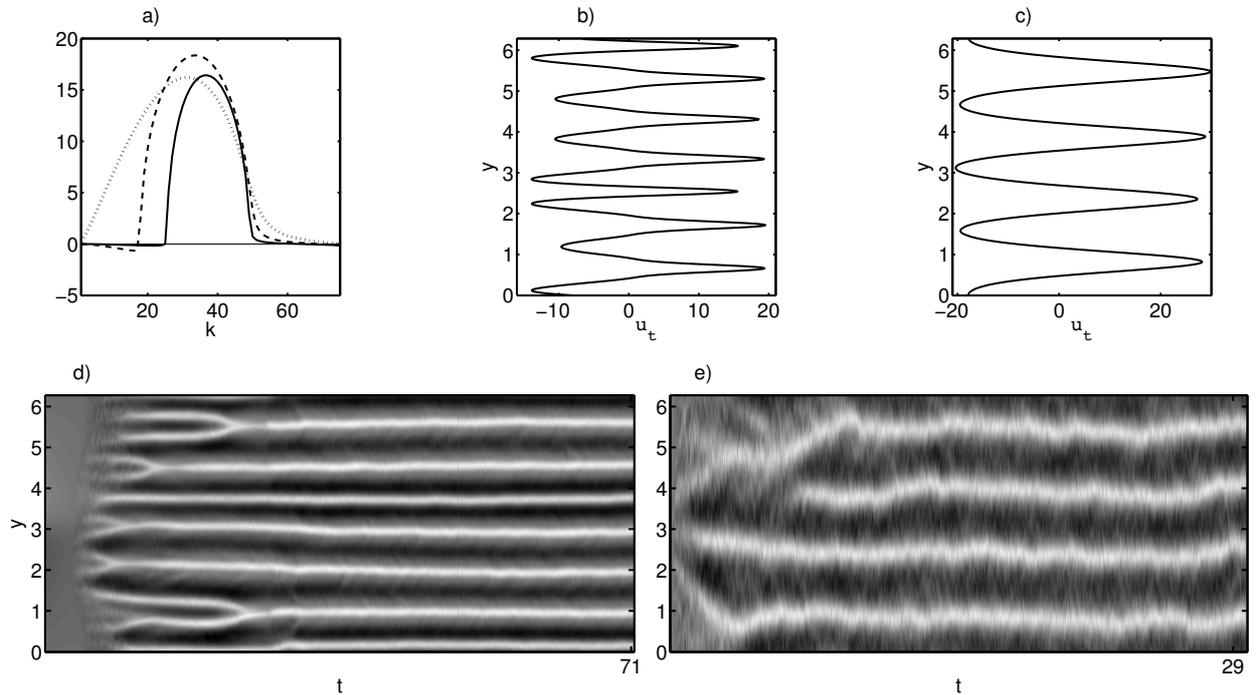}
\caption{a) Growth rates of the most unstable waves as functions of $k_x$ for linear baroclinic instability about the imposed background shear. Weakly supercritical (solid), moderately supercritical (long dash), and strongly supercritical (short dash). The deformation wavenumber is $k_d=50$. b) and c) Time- and zonally-averaged zonal barotropic velocities from the weakly and moderately supercritical reference simulations, respectively; d) and e) time series of the zonally-averaged zonal barotropic velocity for the weakly and moderately supercritical cases, respectively.}\label{fig:QG_BCI}
\end{center}
\end{figure}

When $k_\beta^2<k_d^2$ the imposed background shear is linearly unstable to Rossby waves of the form $q_j = \hat{q}_j$exp$\{i(k_x x +k_y y - c t)\}$ \citep[see, e.g.\,ref.][ch. 6]{Vallis06}.
The most unstable modes occur for $k_y=0$; when $k_\beta \ll k_d$ the unstable range is approximately $|k_x|\in(k_\beta/\sqrt{2},k_d)$, though modes with $|k_x|\ge k_d$ are slightly destabilized by bottom friction, with peak instability at $|k_x|\approx 0.6k_d$.
In the absence of bottom friction the system is marginally stable when $k_\beta^2=k_d^2$.
We consider three parameter settings: weakly supercritical ($k_\beta^2 = k_d^2/2$, $r=1$), moderately supercritical ($k_\beta^2 = k_d^2/4$, $r=4$) and strongly supercritical ($k_\beta=0$, $r=16$); in GM these parameter regimes are referred to as `low latitude' (weakly supercritical), `mid latitude' (moderately supercritical), and `high latitude' (strongly supercritical) because $k_\beta$ decreases as latitude increases.
The value of the bottom friction coefficient $r$ is increased with supercriticality in order to keep the inverse energy cascade from reaching the size of the box.
Growth rates of the unstable modes with $k_y=0$ for the three model configurations are shown as functions of $k_x$ in Fig.~\ref{fig:QG_BCI}a.
Although the maximum growth rate is similar in each of the three parameter settings, the range of unstable wavenumbers grows from the weakly unstable regime to the strongly unstable one.

Figure \ref{fig:DNS} shows snapshots of the barotropic potential vorticity $q_t$ from the three reference simulations.
The weakly and moderately supercritical dynamics organize into seven and four zonal jets, respectively, with vortical eddies, filaments, and waves superimposed, and the strongly supercritical dynamics organize into a sea of vortices and filaments of various sizes.
Figure \ref{fig:QG_BCI}b-d shows time-series and a time-average of the zonally-averaged barotropic zonal velocity for the weakly and moderately supercritical cases.
The jets in both the weakly and moderately supercritical cases are asymmetric with stronger eastward velocity.
The jets in the moderately supercritical case are less pronounced than in the weakly supercritical case, in the sense that their signature in the barotropic vorticity is weaker, and they are more difficult to correctly reproduce in a coarse-resolution simulation.
The four jets in figure \ref{fig:QG_BCI}c have maximum velocity approximately $30$; for comparison, the RMS barotropic velocity for this simulation is $160$, so the jets are an important, but not dominant feature.

The emergence of jets at the large scales of quasigeostrophic turbulence is a well-studied subject, reviewed recently by \citet{DM08}.
Although the jets affect and are affected by the small scales, their formation and maintenance is largely independent of the details of this interaction: jets form on the large scales as long as the small scales provide an upscale cascade of energy.
This is emphasized by our experiments, described below, using coarse-resolution models that represent the small scales by stochastic terms completely uncorrelated from the large scales (the `uncorrelated closure' in section \ref{sec:ThreeClosures}).
The emergence of jets at the large scales is only one of the statistics used here to measure the performance of stochastic superparameterization.

The dynamics generate a meridional heat flux proportional to the domain-integral of $v_t\psi_c$ which acts to erode the background temperature gradient associated with the imposed mean shear.
This heat flux depends strongly on the strength of bottom friction $r$ and on $k_\beta$; thorough investigations of the parameter space are provided by \citet{TY06,TY07}.
The zonal jets that appear in the weakly and moderately supercritical simulations act as barriers to the meridional transport of heat so that the heat flux varies by over two orders of magnitude between the three test cases: the time-averaged, domain-integrated values of $v_t\psi_c$ are $1.03$, $23.3$, and $207$ for the weakly, moderately, and strongly supercritical cases, respectively.
The heat flux for all simulations is reported in table \ref{table}.
Figure \ref{fig:Diagnostics}d shows the one-dimensional heat flux cross-spectrum from each experiment, i.e.\,the time- and angle-averaged value of $\hat{v}_t^*\hat{\psi}_c$ where the hat represents the Fourier transform and $^*$ represents the complex conjugate.
The values are scaled to unit amplitude and offset such that the spectrum for the weakly supercritical case occupies the interval $[0,1]$ on the ordinate axis, moderately supercritical occupies $[1,2]$, and strongly supercritical occupies $[2,3]$.
The heat flux in the moderately and strongly supercritical simulations is generated primarily by wavenumbers with $k\le10$, but in the weakly supercritical simulation the peak of the heat flux spectrum is at $k>10$.
This shows that the heat flux is a difficult statistic to correctly predict in a coarse resolution simulation at weak supercriticality.

The energy flow for this paradigm model of geophysical turbulence is discussed by \citet{Salmon98}.
Kinetic energy cascades upscale from the unstable modes near the deformation scale to a halting scale determined by bottom friction and $k_\beta$, where it becomes primarily barotropic and is dissipated through bottom friction.
At the large scales, the barotropic meridional velocity interacts with the imposed baroclinic potential vorticity gradient (the terms $k_d^2 v_i$ in equation (\ref{eqn:QGPV})) to generate large-scale baroclinic potential vorticity and associated potential energy $k_d^2(\psi_1-\psi_2)^2/2$.
The potential energy generated at large scales cascades downscale towards the deformation radius, where it is converted to barotropic kinetic energy by baroclinic instability.

In the following section we develop a parameterization for coarse-resolution simulations where the coarse grid Nyquist wavenumber is smaller than the deformation wavenumber.
In such simulations, any successful parameterization must both absorb the downscale cascade of potential energy, and more importantly produce an inverse cascade of kinetic energy. 
We develop four eddy closures of increasing sophistication based on a stochastic multiscale model of the eddies.
Three of the closures are stochastic, and are based on a random-orientation, reduced-dimensional approximation of the unresolved eddies, while the fourth uses the expected value of one of the stochastic closures and is therefore deterministic rather than stochastic.
All four closures are able to absorb the downscale potential energy cascade and generate the inverse energy cascade.

\section{\label{sec:Formulation}Formulation of Stochastic Superparameterization}
In this section we develop the mathematical framework underpinning stochastic superparameterization, and formulate four closures, where the eddy terms are either deterministic or stochastic and either nonlinearly dependent on the local mean flow or independent of it.
The stochastic closures are based on reduced dimensional embedded domains with random directions.
In the first three closures the eddy energy level is a tunable constant in space and time. 
In the fourth closure the eddy terms are nonlinear and stochastic, and the eddy energy level varies in space and time as the solution of a heuristically-motivated energy equation.
The stochastic, nonlinear closures are the main focus, with the uncorrelated closure included as a `null hypothesis' and the deterministic closure included primarily for completeness.
Results of simulations using the four closures are presented in section \ref{sec:Results}.

\subsection{Point Approximation\label{sec:PA}}
Following GM we develop a multiscale equation set as follows.
First, apply a Reynolds average $\ov{(\cdot)}$ to the governing equations (\ref{eqn:QGPV}) to generate `mean' equations
\begin{gather}\notag
\pd{t}\oq_1=-\nabla\cdot(\ov{\bu_1q_1})-\pd{x}\oq_1-(k_\beta^2 + k_d^2)\ov{v}_1-\nu\nabla^8\oq_1,\\\notag
\pd{t}\oq_2=-\nabla\cdot(\ov{\bu_2q_2})+\pd{x}\oq_2-(k_\beta^2 - k_d^2)\ov{v}_2-r\nabla^2\opsi_2-\nu\nabla^8\oq_2\\\notag
\oq_1 = \nabla^2\opsi_1+\frac{k_d^2}{2}(\opsi_2-\opsi_1),\\
\oq_2 = \nabla^2\opsi_2-\frac{k_d^2}{2}(\opsi_2-\opsi_1).\label{eqn:QGPV_Mean}
\end{gather}
Subtracting from the original gives equations for the `eddy' part of the flow
\begin{gather}\notag
\pd{t}q_1'=-\nabla\cdot(\bu_1q_1)'-\pd{x}q_1'-(k_\beta^2 + k_d^2)v_1'-\nu\nabla^8q_1',\\\notag
\pd{t}q_2'=-\nabla\cdot(\bu_2q_2)'+\pd{x}q_2'-(k_\beta^2 - k_d^2)v_2'-r\nabla^2\psi_2'-\nu\nabla^8q_2'\\\notag
q_1' = \nabla^2\psi_1'+\frac{k_d^2}{2}(\psi_2'-\psi_1'),\\
q_2' = \nabla^2\psi_2'-\frac{k_d^2}{2}(\psi_2'-\psi_1').
\end{gather}
The mean and eddies are coupled through the potential vorticity flux
\begin{gather*}
\ov{\bu_jq_j} = \bbu_j\oq_j+\ov{\bu_j'q_j'}\\
(\bu_jq_j)' = \bu_jq_j - \ov{\bu_j q_j}.
\end{gather*}
We write the eddy potential vorticity flux without approximation as
\begin{equation}
\nabla\cdot(\ov{\bu_j'q_j'}) = \frac{k_d^2(-1)^j}{2} \nabla\cdot(\ov{\bu_j'(\psi_1'-\psi_2')})
+\left(\pd{x}^2-\pd{y}^2\right)\ov{u_j'v_j'}+ \pd{xy}\left(\ov{(v_j')^2}-\ov{(u_j')^2}\right)
\end{equation}
where the first term is a `heat' or `buoyancy' flux divergence and the remaining terms are the curl of the divergence of the Reynolds stress.
This expansion of the potential vorticity flux is sometimes called the `Taylor identity' after \citet{T15}; for further discussion see \citep{DM08}.

We develop a multiscale formulation by applying the `point approximation' which imposes a dynamical scale separation through the use of embedded domains.
Specifically, we introduce new coordinates $q_j' = q_j'(\tilde{x},\tilde{y},\tau;x,y,t)$ and interpret all derivatives acting on eddy variables in the eddy equation as derivatives in the new coordinates, e.g.
\begin{equation*}
\pd{t}q_j' \to \pd{\tau} q_j',\;\;\pd{x}\psi_j'\to\pd{\tilde{x}}\psi_j'.
\end{equation*}
Thus, at each point $(x,y,t)$ of the physical domain there is an embedded domain with coordinates $(\tilde{x},\tilde{y},\tau)$.
The mean variables do not depend on the new coordinates.
The eddy equations become
\begin{gather}\notag
\pd{\tau}q_1'=-\tnabla\cdot(\bu_1'q_1') -\bbU_1\cdot\tnabla q_1'-\bu_1'\cdot\nabla\ov{Q}_1-\nu\tnabla^8q_1',\\\notag
\pd{\tau}q_2'=-\tnabla\cdot(\bu_2'q_2') - \bbU_2\cdot\tnabla q_2'-\bu_2'\cdot\nabla\ov{Q}_2-r\tnabla^2\psi_2'-\nu\tnabla^8q_2'\\\notag
q_1' = \tnabla^2\psi_1'+\frac{k_d^2}{2}(\psi_2'-\psi_1'),\\
q_2' = \tnabla^2\psi_2'-\frac{k_d^2}{2}(\psi_2'-\psi_1')\label{eqn:PA}
\end{gather}
where $\tnabla = (\pd{\tilde{x}},\pd{\tilde{y}})$, $\bbU_j=\bbu_j-(-1)^j\bm{\hat{x}}$ and $\ov{Q}_j = \oq_j+(k_\beta^2-(-1)^jk_d^2)y$.
Consistent with this introduction of new coordinates we re-interpret the Reynolds average as an average over the new coordinates.

The point approximation is similar to multiple-scales asymptotics (e.g.\,\citep{GSM12,MPP85}) in that the eddies evolve on new independent coordinates where the mean variables are constant.
The primary difference is that multiple-scales asymptotics assumes that the large-scale derivatives are asymptotically small compared to the small-scale derivatives; loosely, $\pd{x}\ll\pd{\tilde{x}}$.
If the mean variables evolve only on scales much larger than the eddies, then the point approximation will recover $\pd{x}\ll\pd{\tilde{x}}$, similar to the asymptotic method.
If, on the other hand, the mean variables and the eddies both display variation on scales close to the coarse grid scale then the assumption of scale separation will be false, casting doubt on the results of the asymptotic analysis.

The point approximation is fairly severe from the perspective of scales immediately above and below the scale of the large-scale computational grid: the approximation causes the scales immediately above the grid scale to appear constant in comparison with those immediately below the grid scale. 
But for scales well separated from the coarse grid scale the approximation improves, and becomes more similar to a multiple-scales asymptotic approximation.
Note that it is possible to make the point approximation in one coordinate at a time by allowing, for example, the mean and eddies to share the $t$ coordinate, or the $y$ coordinate.
In the present situation the mean and eddies share the vertical coordinate, which is discretized into two layers.

There is some ambiguity in applying the point approximation to the mean equation, since the interpretation of the eddy terms is not unique.
For example, one might choose to interpret the divergence of the eddy potential vorticity flux as 
\begin{equation}
\ov{\bu_1'q_1'} = \ov{\tnabla^\bot\psi_1' \tnabla^2\psi_1'}+\frac{k_d^2}{2}\ov{\bu_1'(\psi_2'-\psi_1')}= \frac{k_d^2}{2}\ov{\bu_1'\psi_2'}
\end{equation}
thus ignoring the Reynolds stresses.
In cases that allow such ambiguity, guidance can be provided by physical intuition and mathematical analysis.
As an example of the former, in the phenomenology of turbulence described above (\citep{Salmon98}) the inverse cascade of kinetic energy operates primarily in the barotropic mode, whereas, if the Reynolds stresses were ignored there would be no eddy terms in the barotropic equation.
As an example of the latter, the asymptotic analysis of \citet{GSM12} demonstrates that the Reynolds stresses should not be neglected.

The point approximation provides a means of formally deriving equations governing the large and small scales (and their coupling) that is different from the framework of \citet{Grabowski04}.
A particular benefit of the point approximation is that it allows the horizontal gradient of mean variables to appear in the eddy equations, e.g.\,via the term $\bu_j'\cdot\nabla\ov{Q}_j$ in equation (\ref{eqn:PA}).
Terms of this sort are of fundamental importance; they allow, for example, the growth rate of baroclinic instability to vary depending on latitude (through $k_\beta$).

Although the point approximation provides a basis for a conventional SP simulation, where the eddy equations are solved on horizontally periodic domains embedded in the coarse computational grid, such simulations in this context would be as expensive or even more costly than direct simulation.
In other settings \citep[e.g.][]{GS99,XMG09,CHJM11} computational savings have been achieved by reducing the dimensionality of the embedded domains, for example by making the eddy variables depend only on $\tilde{x}$ and $\tau$ but not $\tilde{y}$.
In the current context this strategy causes the eddy equations to linearize since the nonlinear term has the form $\tnabla\cdot(\bu_i'q_i')=\pd{\tilde{x}}\psi_i'\pd{\tilde{y}}q_i'-\pd{\tilde{y}}\psi_i'\pd{\tilde{x}}q_i'$, which reduces to zero when the eddy variables vary in only one spatial coordinate; this effect is desirable from the standpoint of computational efficiency, but not from the standpoint of realism since in reality the small-scale eddies are strongly nonlinear and turbulent.

Furthermore, in conventional SP simulations the embedded domains are usually given a size smaller than or equal to the large-scale grid; the result is that there is a scale gap of at least a factor of two between the Nyquist wavenumber of the large-scale grid and the smallest wavenumber of the embedded domains.
This can cause an SP simulation to completely miss an important range of unstable wavenumbers.
For example, in the weakly supercritical reference case the band of unstable wavenumbers is approximately $k_x\in[25,50]$ (see figure \ref{fig:QG_BCI}a). 
For an SP simulation with a coarse grid Nyquist wavenumber of $25$, and embedded domains that completely fill the grid, the baroclinic instability that drives the system would be completely unresolved because the smallest wavenumber on the embedded domains would be $50$.
The scale gap can be lessened by increasing the size of the embedded domains, but this further increases the computational expense and does not robustly solve the problem of limited-wavenumber instabilities: if the embedded domains in the above example are made twice the size of the coarse grid then they resolve wavenumbers $25$, $50$, $75$, etc.\,and still miss the instability.

\subsection{Conditional Gaussian Closure\label{sec:GC}}
To alleviate the aforementioned difficulties associated with conventional SP we apply a Gaussian closure for the eddies wherein we approximate the nonlinearity in the eddy equations by additive stochastic forcing and linear deterministic damping
\begin{gather}\notag
\pd{\tau}q_1'=F_1'-\Gamma q_1'-\bbU_1\cdot\tnabla q_1'-\bu_1'\cdot\nabla\ov{Q}_1-\nu\tnabla^8q_1',\\
\pd{\tau}q_2'=F_2'-\Gamma q_2'-\bbU_2\cdot\tnabla q_2'-\bu_2'\cdot\nabla\ov{Q}_2-r\tnabla^2\psi_2'-\nu\tnabla^8q_2'.\label{eqn:GC}
\end{gather}
The forcing terms $F_i'$ are spatially correlated and white in time and $\Gamma$ is a positive-definite pseudo-differential operator representing turbulent damping (further details below).
We emphasize that the stochastic approximation adopted above is an approximation of strongly nonlinear, turbulent dynamics; weakly nonlinear, temporally chaotic but non-turbulent eddy behavior of the type seen in transition-to-turbulence scenarios would require a different kind of stochastic model.

Similar approximations, wherein the eddy-eddy nonlinearity is replaced by a Gaussian stochastic forcing and damping, have often been made in the context of quasigeostrophic turbulence (see, e.g.\,\citep{DelSole04,FI09} and references therein).
The CE2 (second order cumulant expansion) method used by \citet{TM13} and \citet{SY12} drops the eddy-eddy nonlinearity altogether, without replacing it by a stochastic approximation.
\citet{SM13} develop a modified quasilinear Gaussian approximation that includes a more sophisticated, energy-conserving approximation of the eddy-eddy nonlinearity; they also show that the CE2 closure necessarily evolves to an incorrect marginally-stable statistical equilibrium regardless of the external forcing in the Lorenz-96 model \citep{L96,L06}.
A key difference of the current approach, motivated by \citet{MG09}, is that the Gaussian stochastic model is developed here only for the small scales in a multiscale framework based on the point approximation above.

Following GM and \citep{GM14} we model the eddy variables as spatially-homogeneous random functions in a formally infinite domain $\bm{\tilde{x}}=(\tilde{x},\tilde{y})\in\mathbb{R}^2$ with the following spectral representation
\begin{equation}
q_j' = \iint \hat{q}_j e^{i\bm{k}\cdot\bm{\tilde{x}}}\text{d}W_{\bm{k}}
\end{equation}
where $W_{\bm{k}}$ is a complex Weiner process and $\hat{q}_j$ depends on $\bm{k}=(k_x,k_y)$.
The use of formally infinite embedded domains instead of periodic ones is convenient since it allows a continuum of possible eddy scales, thereby avoiding the difficulty associated with conventional SP of missing instabilities that occur on a limited range of wavenumbers.

The Fourier transform of the eddy equations is
\begin{gather}\notag
\frac{\text{d}}{\text{d}\tau}\hat{q}_1=- i(\bbU_1\cdot\bm{k})\hat{q}_1 - (i\bm{k}\times\nabla\ov{Q}_1)\hat{\psi}_1 +A_{1,\bm{k}}\dot{W}_{1,\bm{k}}-(\gamma_{\bm{k}}+\nu k^8)\hat{q}_1,\\\notag
\frac{\text{d}}{\text{d}\tau}\hat{q}_2=- i(\bbU_2\cdot\bm{k})\hat{q}_2 - (i\bm{k}\times\nabla\ov{Q}_2)\hat{\psi}_2 + A_{2,\bm{k}}\dot{W}_{2,\bm{k}}+rk^2\hat{\psi}_2-(\gamma_{\bm{k}}+\nu k^8)\hat{q}_2,\\\notag
\hat{q}_1 = -k^2\hat{\psi}_1 + \frac{k_d^2}{2}(\hat{\psi}_2-\hat{\psi}_1),\\
\hat{q}_2 = -k^2\hat{\psi}_2 + \frac{k_d^2}{2}(\hat{\psi}_1-\hat{\psi}_2)
\end{gather}
where $k = |\bm{k}|$, $W_{j,\bm{k}}$ are independent complex Weiner processes $\mathbb{E}[W_{j,\bm{k}}W_{i,\bm{k}'}]=\delta_{ij}\delta_{\bm{k}\bm{k}'}$, and $A_{j,\bm{k}}$ are complex constants.
We write this as a system for $\hat{\psi}_1$ and $\hat{\psi}_2$
\begin{equation}
\text{d}\left(\begin{array}{c} \hat{\psi}_1 \\ \hat{\psi}_2 \end{array}\right) =\bL_{\bm{k}}
\left(\begin{array}{c} \hat{\psi}_1 \\ \hat{\psi}_2 \end{array}\right) \text{d}\tau + \sigma_{\bm{k}}\text{d}\bm{W}_{\bm{k}}
\end{equation}
where $\sigma_{\bm{k}}$ is a complex-valued matrix, $\bm{W}_{\bm{k}}$ is a vector of independent complex Weiner processes, and 
\begin{equation}
\bL_{\bm{k}}= -(\gamma_k+\nu k^8)\text{\bf I}+\text{\bf Q}_k^{-1}\left(-i\left[
\begin{array}{c c}\bbU_1\cdot\bm{k} & 0\\0 & \bbU_2\cdot\bm{k}\end{array}\right]\text{\bf Q}_k+\left[\begin{array}{c c}
-i\bm{k}\times\nabla\ov{Q}_1 & 0\\
0 & r k^2-i\bm{k}\times\nabla\ov{Q}_2\end{array}\right]\right),
\end{equation}
\begin{equation}
\text{\bf Q}_k = \left[\begin{array}{c c} -\left(\frac{k_d^2}{2}+k^2\right) & \frac{k_d^2}{2}\\ \frac{k_d^2}{2} & -\left(\frac{k_d^2}{2}+k^2\right)\end{array}\right].
\end{equation}

At this point note that the eddy terms that appear in the mean equation, e.g.\, $\ov{\bu_1'\psi_2'}$, all consist of the average over $\tilde{\bm{x}}$ and $\tau$ of quadratic products of eddy variables.
The spatial average of a quadratic product is related to an integral over the Fourier coefficients by the Plancherel theorem
\begin{equation}
\iint f g \text{d}\bm{\tilde{x}} = \iint \hat{f}^*\hat{g}\text{d}\bm{k}
\end{equation}
where $^*$ denotes the complex conjugate (and conjugate transpose for vectors and matrices).
Furthermore, because the eddies are spatially homogeneous, the spatial average is equal to an ensemble average.
Thus terms in the eddy potential vorticity flux can be calculated as, e.g.
\begin{align}\notag
\ov{u_1'(\psi_2'-\psi_1')}=\ov{u_1'\psi_2'}&=i\iint\left(\eps\int_0^{\eps^{-1}}\mathbb{E}\left[k_y\hat{\psi}_1^*\hat{\psi}_2\right]\text{d}\tau\right)\text{d}\bm{k}\\
&=\iint k_y\left(\eps\int_0^{\eps^{-1}}\mathbb{E}\left[\mathcal{I}\{\hat{\psi}_1\hat{\psi}_2^*\}\right]\text{d}\tau\right)\text{d}\bm{k}\label{eqn:ub}\\
\ov{u_i'v_i'} &= \iint k_xk_y \left(\eps\int_0^{\eps^{-1}}\mathbb{E}\left[|\hat{\psi}_{i}|^2\right]\text{d}\tau\right)\text{d}\bm{k}\label{eqn:uv}
\end{align}
where $\mathbb{E}$ denotes an ensemble average, $\mathcal{I}\{\cdot\}$ denotes the imaginary part of a complex number, and the average over $\tau$ has length $\eps^{-1}$.
The formulas of the remaining terms are found in \ref{sec:Appendix}.
This suggests that, rather than simulate solutions of the linear stochastic eddy PDE directly, we instead solve for the evolution of the quadratic products involved in the eddy potential vorticity flux.

The covariance of the Fourier coefficients of the eddy streamfunction
\begin{equation}
C_{\bm{k}} = \mathbb{E}\left[\begin{array}{c c}|\hat{\psi}_1|^2 & \hat{\psi}_1\hat{\psi}_2^*\\\hat{\psi}_1^*\hat{\psi}_2&|\hat{\psi}_2|^2\end{array}\right]
\end{equation}
evolves according to the linear, autonomous ordinary differential equation
\begin{equation}
\frac{\text{d}}{\text{d}\tau}C_{\bm{k}} = \bL_{\bm{k}}C_k + C_k \bL_{\bm{k}}^* + \sigma_{\bm{k}}\sigma_{\bm{k}}^*\label{eqn:Ck}
\end{equation}
which is obtained from the It\=o formula.
Note that $\bL_{\bm{k}}$ depends on the mean variables and their derivatives and thus provides coupling to the large scales.
To evolve $C_k$ according to this equation one must specify an initial condition, $\gamma_{\bm{k}}$, and $\sigma_{\bm{k}}\sigma_{\bm{k}}^*$; that is, one must specify $C_{\bm{k}}(\tau=0)$ and the forcing and damping that model the nonlinear eddy-eddy interaction.

We deal first with the details of the forcing and damping.
Following GM we specify $\gamma_{\bm{k}}$ and $\sigma_{\bm{k}}\sigma_{\bm{k}}^*$ by requiring the solution $C_{\bm{k}}$ to relax towards a stable equilibrium with phenomenological properties in the absence of mean variables, i.e.\,when $\bbU_j=\nabla\ov{Q}_j=0$.
In the following we (i) detail the properties of the equilibrium covariance (equations (\ref{eqn:CkEq}) and (\ref{eqn:n_k})), (ii) specify under what conditions the system should approach this equilibrium (equation (\ref{eqn:OU})), and finally (iii) provide remaining assumptions to complete the specification of $\gamma_{\bm{k}}$ and $\sigma_{\bm{k}}\sigma_{\bm{k}}^*$ (equation (\ref{eqn:gamma0})).
We specify the equilibrium covariance such that
\begin{enumerate}
\item it is isotropic (a function only of $k = |\bm{k}|$)
\item the energy spectrum is proportional to $k^{-5/3}$ for $k<k_d$ and to $k^{-3}$ for $k\ge k_d$
\item the barotropic kinetic energy equals the baroclinic energy at every $\bm{k}$
\item $\alpha\mathbb{E}[|\hat{\psi}_1|^2] = \mathbb{E}[|\hat{\psi}_2|^2]$ with $\alpha>0$
\item $\mathbb{E}\left[\mathcal{I}\{\hat{\psi}_1\hat{\psi}_2^*\}\right]=0$.
\end{enumerate}

The first property guarantees that the equilibrium spectrum, and hence the stochastic approximation of the nonlinear term, does not bias the Reynolds stresses since an isotropic spectrum produces $\ov{u_i'v_i'}=0$ and $\ov{(u_i')^2} = \ov{(v_i')^2}$.
Quasigeostrophic turbulence on a $\beta$-plane (i.e. $k_\beta\neq0$) is known to develop an anisotropic `dumbell' spectrum, but this affects primarily the large scales, so isotropy remains an appropriate assumption for the small scales.
\begin{figure}
\begin{center}
\includegraphics[width=\textwidth]{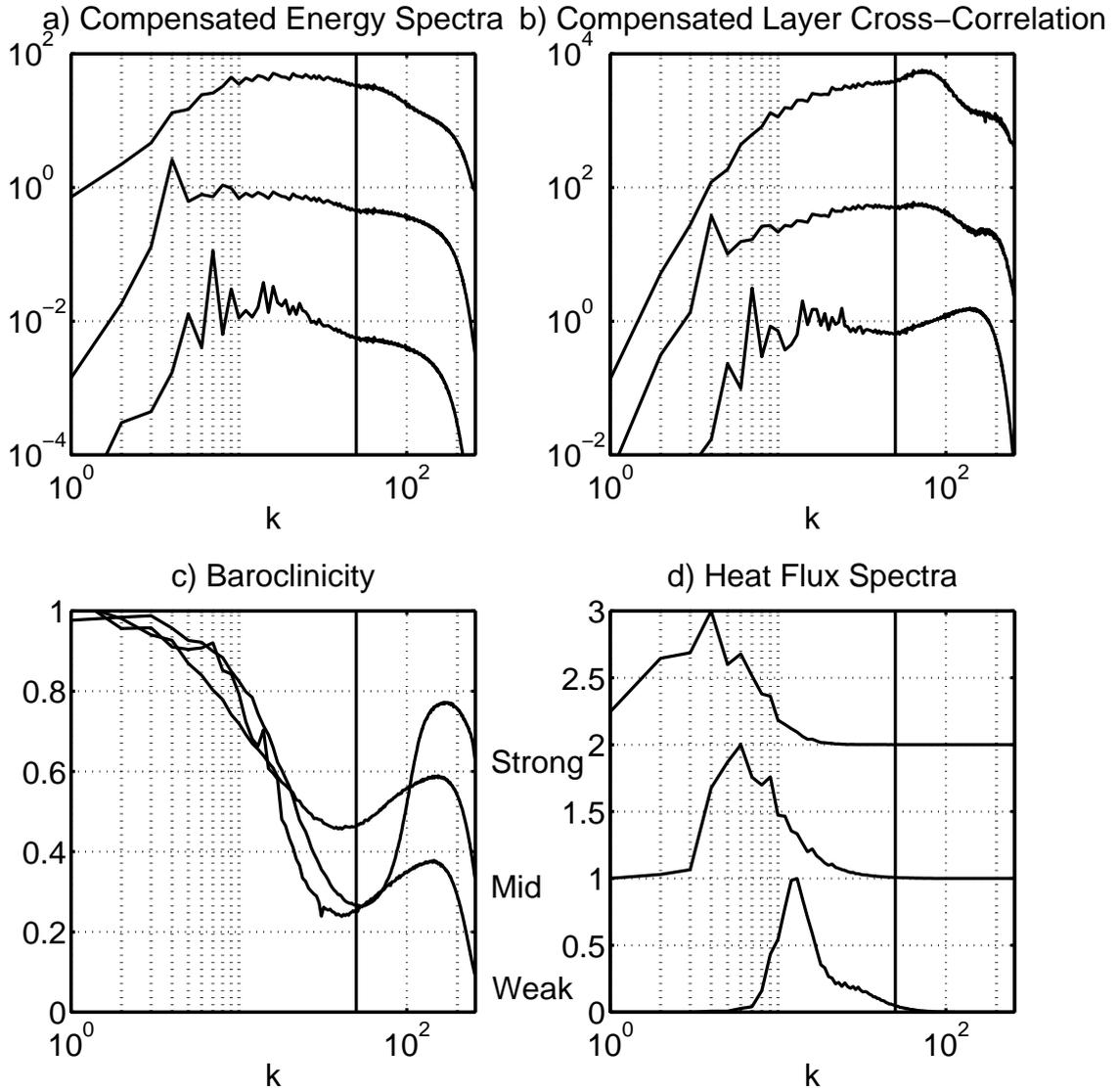}
\caption{a) Time- and angle-averaged energy spectra from the reference simulations compensated by $k^{-5/3}$ for $k<k_d$ and by $k=k_d^{4/3}k^{-3}$ for $k\ge k_d$; in a), b) and d) the uppermost line is the strongly supercritical case, the middle line is the moderately supercritical case, and the lower line is the weakly supercritical case. b) Time and angle averaged $\mathcal{R}\{\hat{\psi}_1\hat{\psi}_2^*\}$ compensated by the value from the equilibrium covariance. c) Time- and angle-averaged kinetic energy spectrum in the lower layer divided by the upper layer spectrum; the legend is to the right of the plot. d) Time- and angle-averaged heat flux cross-spectrum $\hat{v}_t^*\hat{\psi}_c$ normalized to unit amplitude and offset; the goal is primarily to show that the peak of the heat flux spectrum lies at $k>10$ for the weakly supercritical case, at $k=6$ for the moderately supercritical case, and at $k=4$ for the strongly supercritical case. The deformation wavenumber $k_d=50$ is shown as a vertical line in each panel.}\label{fig:Diagnostics}
\end{center}
\end{figure}

The second property, the slope of the energy spectrum, is well known from the phenomenology of quasigeostrophic turbulence.
In figure \ref{fig:Diagnostics}a we plot the time- and angle-averaged energy spectra from the three reference simulations compensated by the energy spectrum of the equilibrium covariance. 
Each compensated spectrum is approximately flat (indicating approximately correct spectral slope) for $k>10$, and falls off due to viscosity at small scales.
Although GM included an exponential decrease in the equilibrium energy spectrum at small scales, we leave the small-scale spectrum at $k^{-3}$ to demonstrate that the results are not sensitive to the small-scale properties of the equilibrium spectrum.

GM specified that the total kinetic energy in the equilibrium equals twice the available potential energy at each $k<k_d$
\begin{equation}
k^2(|\hat{\psi}_1|^2+|\hat{\psi}_2|^2)=k_d^2(|\hat{\psi}_1|^2+|\hat{\psi}_2|^2 - 2\mathcal{R}\{\hat{\psi}_1\hat{\psi}_2^*\}).
\end{equation}
We instead specify the equilibrium to have barotropic kinetic energy equal to baroclinic energy (potential energy plus baroclinic kinetic energy) at every $k$, 
\begin{equation}
k^2(|\hat{\psi}_1|^2+|\hat{\psi}_2|^2+2\mathcal{R}\{\hat{\psi}_1\hat{\psi}_2^*\})=(k^2+k_d^2)(|\hat{\psi}_1|^2+|\hat{\psi}_2|^2 - 2\mathcal{R}\{\hat{\psi}_1\hat{\psi}_2^*\}).
\end{equation}
The former option implies that $\mathcal{R}\{\hat{\psi}_1\hat{\psi}_2^*\}\to0$ as $k\to k_d$, which is not consistent with the reference simulations.
For comparison, \citet{LH95} suggest that the barotropic kinetic energy should be approximately five times larger than the total baroclinic energy at each wavenumber.
Figure \ref{fig:Diagnostics}b shows the time- and angle-averaged value of  $\mathcal{R}\{\hat{\psi}_1\hat{\psi}_2^*\}$ for each of the three reference simulations compensated by the equilibrium spectrum that assumes equipartition of barotropic and baroclinic energies. 
The compensated spectra are approximately flat in each case, but only to a very loose approximation.
Rather than tune the properties of the equilibrium spectrum to match the reference simulations more closely, we use the simpler assumption of equipartition of barotropic and baroclinic energies (the third property above) to underscore the success of the method even with an imperfect equilibrium.

The fourth property specifies the partition of kinetic energy between layers.
For $\alpha=1$, which was used by GM, the layers have equal energy.
Figure \ref{fig:Diagnostics}c shows the ratio of the time- and angle-averaged $|\hat{\psi}_2|^2$ and $|\hat{\psi}_1|^2$ for each of the three reference simulations.
While the layers have approximately equal energy for large scales, at smaller scales the lower layer has less energy in all of the reference simulations.
This accords with standard quasigeostrophic turbulence theory (which predicts barotropic large scales) and with the expectation of having lower energy in the lower layer due to bottom friction.
We set $\alpha=1/4$ for the weakly supercritical case and $\alpha=1/2$ for the moderately and strongly supercritical cases; some results with $\alpha=1$ are also provided for comparison in section \ref{sec:Results}.
The differences in the results between $\alpha<1$ and $\alpha=1$ are notable, but small; this, and the fact that we do not further tune $\alpha$ or make it a function of $k$ underscore the robustness of the method.

The fifth property guarantees that the equilibrium has no associated eddy buoyancy flux (see equation (\ref{eqn:ub})), so that the stochastic approximation to the nonlinear terms does not bias the eddy buoyancy flux.
This is also consistent with data in the sense that the time- and angle-average of $\mathcal{I}\{\hat{\psi}_1\hat{\psi}_2^*\}$ is zero in the reference simulations (not shown).
The reference simulations are able to generate nonzero net heat flux because they do not have isotropic $\mathcal{I}\{\hat{\psi}_1\hat{\psi}_2^*\}$.

The five properties listed above are sufficient to specify the equilibrium covariance as
\begin{equation}
C_{\bm{k},\text{eq}}=C_{k,\text{eq}} = A(x,y,t) n_k\left[\begin{array}{c c} \frac{2(2k^2+k_d^2)}{1+\alpha} & k_d^2\\k_d^2 &  \frac{2\alpha(2k^2+k_d^2)}{1+\alpha}\end{array}\right],\label{eqn:CkEq}
\end{equation}
where
\begin{equation}
n_k = \left\{\begin{array}{c l} 0 & \text{ for }k<k_0\text{ and }k>k_{\text{max}},\\
(4 k^{14/3}(k^2+k_d^2))^{-1} & \text{ for } k_0\le k<k_d,\\
 k_d^{4/3}(4 k^6(k^2+k_d^2))^{-1} & \text{ for } k_d\le k \le k_{\text{max}}.\end{array}\right.\label{eqn:n_k}
\end{equation}
Large- and small-scale cutoffs $k_0$ and $k_{\text{max}}$ are introduced to prevent the eddies from including unrealistically large or small scales; $k_0$ is set equal to the Nyquist wavenumber of the coarse grid and $k_{\text{max}}=256$ to the Nyquist wavenumber of the reference simulations.
Note that the equilibrium covariance is symmetric positive definite (and thus a covariance matrix) only for a range of $k$ that depends on $\alpha$; for the values of $\alpha$ chosen here the the equilibriumm covariance matrix is well-defined for all $k$ in the eddy wavenumber range $k_0\le k\le k_{\text{max}}$.
The amplitude of the eddies, $A$, is considered to be tunable and constant across the large-scale spatiotemporal domain except as described in section \ref{sec:EnergyMethods} (with simulation results in section \ref{sec:EnergyResults}).

Having thus specified the equilibrium covariance, it remains to specify $\gamma_{\bm{k}}$ and $\sigma_{\bm{k}}\sigma_{\bm{k}}^*$, and the initial condition for the covariance evolution equation (\ref{eqn:Ck}).
In GM these were specified by requiring the covariance evolution equation (\ref{eqn:Ck}) to relax to the equilibrium covariance when the mean variables are zero, i.e.\,when $\bbU_i=\nabla\ov{Q}_i=0$.
We take the simpler approach here of requiring the covariance evolution equation (\ref{eqn:Ck}) to relax to the equilibrium covariance when the mean variables and the viscous and Ekman dissipation are zero, i.e.
\begin{equation}
2\gamma_{\bm{k}}C_{k,\text{eq}} = \sigma_{\bm{k}}\sigma_{\bm{k}}^*.\label{eqn:OU}
\end{equation}
In general, one must take care to specify the entries of $\sigma_{\bm{k}}\sigma_{\bm{k}}^*$ in such a way that it is symmetric and positive definite; the above approach guarantees this provided that the equilibrium covariance is a covariance matrix.
As in GM, we close the system by requiring $\gamma_{\bm{k}}$ to be isotropic and proportional to the nonlinear eddy timescale at each $k$
\begin{equation}
\gamma_{\bm{k}}=\gamma_k = \left\{\begin{array}{c l}\gamma_0 (k/k_d)^{2/3} & \text{ for }k < k_d,\\
\gamma_0 & \text{ for } k \ge k_d.\end{array}\right.\label{eqn:gamma0}
\end{equation}
As in GM, we set $\gamma_0=30$ so that it is slightly more than sufficient to damp the linear instability of the imposed shear.
We emphasize that this choice doesn't guarantee saturation of the eddy statistics, because mean shear (and associated eddy instability) can become much larger than the imposed shear in the course of a simulation.
Note that the choice of $\gamma_{\bm{k}}$ only specifies $\sigma_{\bm{k}}\sigma_{\bm{k}}^*$ and not $\sigma_{\bm{k}}$ so the properties of the stochastic approximation to the nonlinear terms in the eddy PDE (\ref{eqn:GC}) are not completely defined.

The initial condition for the covariance evolution equation (\ref{eqn:Ck}) is naturally taken to be the equilibrium covariance.
One might alternatively set the initial condition to zero, but we have not explored this option.
In conventional SP, the initial condition of the eddies is tracked from one large-scale time step to the next; this could be done in the present context provided that $C_{\bm{k}}$ is tracked at a finite number of points in $\bm{k}$.
As an alternative to tracking $C_{\bm{k}}$ one might reset the shape of $C_{\bm{k}}$ to the equilibrium at each time step, but change the coefficient $A$ to account for local changes in eddy energy.
This alternative is discussed further in section \ref{sec:EnergyMethods}.

The covariance evolution equation (\ref{eqn:Ck}) can be written as a linear vector equation in the form
\begin{equation}
\frac{\text{d}}{\text{d}\tau}\bm{c}_{\bm{k}} = \bM_{\bm{k}}\bm{c}_{\bm{k}} + \Sigma_k\label{eqn:bM}
\end{equation}
where 
\begin{align*}
\bm{c}_{\bm{k}} &= \mathbb{E}\left[(|\hat{\psi}_1|^2, \mathcal{R}\{\hat{\psi}_1\hat{\psi}_2^*\},\mathcal{I}\{\hat{\psi}_1\hat{\psi}_2^*\},|\hat{\psi}_2|^2)\right],
\end{align*}
$\Sigma_k$ is a vector containing the real and imaginary components of the elements of $\sigma_k\sigma_k^*$, and $\bM_{\bm{k}}$ is the linear coefficient matrix.
The form of the linear propagator {\bf M}$_{\bm{k}}$ is listed in \ref{sec:Appendix}.
The time-average of the covariance evolution, assuming that the equilibrium covariance is used as the initial condition, can be evaluated via
\begin{gather}
\!\eps\int_0^{\eps^{-1}}\bm{c}_{\bm{k}}(\tau) \text{d}\tau= \left[\phi_1(\bM_{\bm{k}}/\eps)+\frac{2\gamma_k}{\eps}\phi_2(\bM_{\bm{k}}/\eps)\right]\bm{c}_{k,\text{eq}}\label{eqn:CkIntegrand1}\\\notag
\phi_1(\text{\bf A}) = \text{\bf A}^{-1}\left[e^{\text{\bf A}}-\text{\bf I}\right],\\\notag
\phi_2(\text{\bf A}) = \text{\bf A}^{-2}\left[e^{\text{\bf A}}-\text{\bf I} - \text{\bf A}\right] = \text{\bf A}^{-1}\left[\phi_1(\text{\bf A})-\text{\bf I}\right].
\end{gather}
Note that the above formula for the time average is only valid when $\bM_{\bm{k}}$ is nonsingular.
However, $\bM_{\bm{k}}$ is singular only on a set of measure zero in $\bm{k}$, which does not affect the eddy terms integrated over $\bm{k}$.

The closure for the eddy terms in the mean equations (\ref{eqn:QGPV_Mean}) can be calculated from their definitions by evaluating the time average from (\ref{eqn:CkIntegrand1}) and the Plancherel integral over $\bm{k}$ by a quadrature.
It is convenient to record the integrals defining the eddy terms in polar form with $(k_x,k_y) = k(\cos(\theta),\sin(\theta))$; for example,
\begin{align}
\ov{u_1'(\psi_2'-\psi_1')}&=\int_0^{2\pi}\int_{k_0}^{k_{\text{max}}}k^2\sin(\theta)\left(\eps\int_0^{\eps^{-1}}\mathbb{E}\left[\mathcal{I}\{\hat{\psi}_1\hat{\psi}_2^*\}\right]\text{d}\tau\right)\text{d}k\text{d}\theta.\label{eqn:ubPolar}\\
\ov{u_i'v_i'} &= \frac{1}{2}\int_0^{2\pi}\int_{k_0}^{k_{\text{max}}} k^3\sin(2\theta) \left(\eps\int_0^{\eps^{-1}}\mathbb{E}\left[|\hat{\psi}_i|^2\right]\text{d}\tau\right)\text{d}k\text{d}\theta.\label{eqn:uvPolar}
\end{align}
The polar-form integrals defining the remaining eddy terms are listed in \ref{sec:Appendix}.

We note finally that the length of the time average $\eps^{-1}$ is tied to the length of the coarse grid time step in conventional SP.
This is natural since the state of the eddies is tracked from one coarse grid time step to the next. 
In the current formulation the eddies are re-set to the equilibrium covariance at every time step, and the length of a coarse-grid time step is generally too short to allow meaningful evolution of the eddies away from their artificial initial condition. 
We therefore allow $\eps^{-1}$ to be larger than the coarse grid step size to give the eddies a longer time to react.
Decoupling $\eps$ from the coarse-grid time step introduces a new tunable parameter, which is not ideal.
It would be less arbitrary to set $\eps$ such that the eddy evolution timescale is shorter than or comparable to the shortest decorrelation time of the mean variables that appear in the eddy equation; however, lacking that information $\eps$ is left here as a tunable parameter.

\subsection{Three Closures: Deterministic, Stochastic, and Uncorrelated\label{sec:ThreeClosures}}
The eddy closure is thus far completely specified up to the choice of the eddy amplitude $A$ and the length of time average $\eps^{-1}$.
In this section we develop three closures where $A$ and $\eps$ are tunable constants.
In each of the closures the radial part of the integrals (\ref{eqn:ubPolar})-(\ref{eqn:uvPolar}) is approximated using a trapezoid-rule quadrature with $k_{\text{max}}-k_0 +1$ equispaced nodes in $k$.
The implementation of each of the closures relies on pre-computing the integrals in the radial direction ($k$) as functions of the large-scale variables $\bm{\hat{k}}\cdot\bbU_c$, $\bm{\hat{k}}\times\nabla\ov{\omega}_c$ and $\bm{\hat{k}}\times\nabla\ov{\omega}_t$ where $\bm{\hat{k}}$ is a unit vector in the direction of $\bm{k}$ (these variables are equivalent to $\bm{\hat{k}}\cdot\bbU_c$, $\bm{\hat{k}}\times\nabla\ov{Q}_1$ and $\bm{\hat{k}}\times\nabla\ov{Q}_2$, but are better conditioned for interpolation).
The ranges of large-scale variables over which the eddy terms are pre-computed depend on the test case; for the moderately supercritical case we use $|\bm{\hat{k}}\cdot\bbU_c|\le 3.5$, $|\bm{\hat{k}}\times\nabla\ov{\omega}_c|\le 10^3$ and $|\bm{\hat{k}}\times\nabla\ov{\omega}_t|\le 1.5\times10^4$.
These values are \emph{a posteriori} verified to cover closely the ranges observed in the actual simulations of the moderately supercritical system (limits for other cases are specified in section \ref{sec:Results}).
In every case, the eddy terms are pre-computed on a grid of $101$ equispaced points in each of the three large-scale variables, and linear interpolation is used to evaluate the radial part of the integral.

The first closure (which we call `deterministic') consists in approximating the integrals defining the eddy terms (\ref{eqn:ubPolar})-(\ref{eqn:uvPolar}) and (\ref{eqn:ub2Polar})-(\ref{eqn:vmuPolar}) by a trapezoid-rule quadrature with $40$ equispaced nodes in $\theta$, and using linear interpolation from the $101^3$ precomputed values of the radial part of the integrals.
Precomputing only the radial direction saves considerable computational effort compared to precomputing both directions of the integral, or neither.

\begin{figure}
\begin{center}
\includegraphics[width=\textwidth]{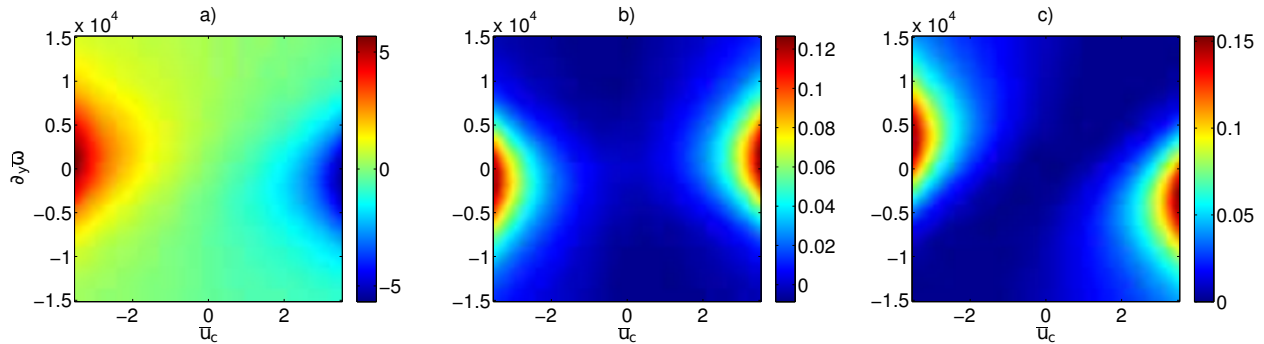}
\caption{Eddy response to baroclinic shear and barotropic vorticity gradient computed using $\eps=25$ and $\gamma_0=30$. a) $(k_d^2/2)\ov{v_1'\psi_2'}$, b) $\ov{(v_1')^2}-\ov{(u_1')^2}$, and c) $\ov{(v_2')^2}-\ov{(u_2')^2}$.}\label{fig:Closure}
\end{center}
\end{figure}
To illustrate the nonlinear behavior of the deterministic closure we display in figure \ref{fig:Closure} the response of the eddy terms to a range of zonal shears with amplitude $|\ov{u}_c|\le 3.5$ and meridional barotropic vorticity gradients with amplitude $|\bm{\hat{k}}\times\nabla\ov{\omega}_t|\le 1.5\times10^4$ (this range is much greater than the background `planetary' vorticity gradient at moderate supercriticality, $k_\beta^2=1250$, but fits the observed range of barotropic vorticity gradients in the simulations); results are displayed for drag coefficient $r=4$, which is the value chosen for the moderately supercritical experiments, and $\eps=25$.
The zonal eddy heat flux and the off-diagonal Reynolds stress $\ov{u_j'v_j'}$ are zero, so figure \ref{fig:Closure} shows the meridional heat flux $(k_d^2/2)\ov{v_1'\psi_2'}$ in panel (a), $\ov{(v_1')^2}-\ov{(u_1')^2}$ in panel (b), and $\ov{(v_2')^2}-\ov{(u_2')^2}$ in panel (c).
The nonlinearity of the closure as a function of the local mean is clearly apparent; even in the absence of a mean meridional vorticity gradient (a horizontal line through the middle of each panel) the eddy terms are not linear functions of the mean zonal baroclinic shear.

However, the inverse cascade of QG turbulence is not well approximated by a deterministic process, since the unresolved scales are only correlated with, and not completely determined by the large scales.
Furthermore, modeling the unresolved eddies as homogeneous random functions in formally infinite embedded domains belies the fact that the subgrid scales at a given location are generally quite inhomogeneous, and consist not of an infinite population of eddies but of a few non-axisymmetric vortices, vortex dipoles, filaments, etc.

This small-scale inhomogeneity can be incorporated into the existing framework by approximating the integrals over $\bm{k}$ that define the eddy terms by a `random' quadrature where the quadrature nodes are randomly chosen.
Motivated by the common practice of using reduced-dimensional embedded domains in conventional SP, and by the success of algorithms based on random plane waves in modelling turbulent diffusion \citep{EM95,EM96,MK99}, we develop a stochastic closure based on approximating the integrals defining the eddy terms (\ref{eqn:ubPolar})-(\ref{eqn:uvPolar}) and (\ref{eqn:ub2Polar})-(\ref{eqn:vmuPolar}) by integrating along one randomly-chosen value of $\theta$.
Specifically, the closure is defined by using a two-point ($\theta$ and $-\theta$) trapezoid-rule quadrature of (\ref{eqn:ubPolar})-(\ref{eqn:uvPolar}) and (\ref{eqn:ub2Polar})-(\ref{eqn:vmuPolar}) where a different value of $\theta$ is chosen at each coarse grid point and at each time step from a uniform distribution in $[0,\pi)$.
The integral in the radial direction is evaluated using linear interpolation from the pre-computed values.
By choosing the direction $\theta$ from a uniform distribution we guarantee that the expected value of the random integral equals the full integral that defines the deterministic closure.
We refer to this second closure as the `correlated stochastic plane wave' closure.

Finally, for comparison, we evaluate the closure based on the random quadrature above, but only sampling the un-evolved equilibrium covariance, i.e.\,the limit $\eps\to\infty$. 
In this limit the eddies become completely uncoupled from the mean and have zero buoyancy flux, and therefore amount to a sophisticated form of structured random-noise forcing.
This closure is included for the sake of demonstrating that, for the weakly and strongly supercritical cases, the large-scale dynamics are fairly insensitive to the precise structure of the small-scale forcing that generates the inverse cascade of kinetic energy.
This third closure is referred to as the `uncorrelated stochastic plane wave' closure.

We find that the stochastic closures developed in this section are made more stable (and make more sense from the standpoint of numerical analysis) by holding the randomly-chosen angle $\theta$ constant through all stages of a single Runge-Kutta time step, instead of changing it at each substage.
It is possible to incorporate a finite decorrelation time into the angle by modeling it as a random walk, which keeps the expected value of the stochastic closures equal to the deterministic closure, but we have not pursued this option.
However, tying the decorrelation time of the angle directly to the coarse-grid time step, as we have chosen here, makes the algorithm sensitive to the size of that time step.
In all the tests of the stochastic closures reported in section \ref{sec:Results} we keep the time step fixed at $2\times10^{-4}$, which is more than adequate to resolve the deterministic part of the dynamics.

\subsection{A Fourth Closure Based on Energy Conservation\label{sec:EnergyMethods}}
The foregoing closures use the equilibrium covariance as an initial condition for the eddies at every time step, and keep the eddy energy density (proportional to $A$) constant in space and time.
In reality, eddy energy is continually exchanged with the large scales in such a way that the total energy is conserved (absent forcing and dissipation).
The energy equation for the mean is obtained by multiplying the mean equation for layer $j$ by $-\opsi_j$, summing the layers, and integrating over $\bm{x}$; the result is
\begin{multline} 
\frac{1}{2}\frac{\text{d}}{\text{d}t}\iint |\bbu_1|^2+|\bbu_2|^2+\frac{k_d^2}{2}(\opsi_1-\opsi_2)^2\text{d}\bm{x}=- k_d^2\iint\left(\ov{v}_1\opsi_2+\ov{\bu_1'\psi_2'}\cdot\nabla\opsi_c\right)\text{d}\bm{x}\\
-D+ \iint\sum_i\left[\ov{u_i'v_i'}(\pd{x}^2-\pd{y}^2)\opsi_i + \left(\ov{(v_i')^2}-\ov{(u_i')^2}\right)\pd{xy}\opsi_i\right]\text{d}\bm{x}
\end{multline}
where $-D$ denotes frictional dissipation and $-k_d^2\ov{v}_1\opsi_2$ represents energy gained through interaction with the imposed background shear.

The energy in the initial condition for the eddies is 
\begin{align}\notag
\frac{1}{2}\iint |\bu_1'|^2+|\bu_2'|^2+\frac{k_d^2}{2}(\psi_1'-\psi_2')^2\text{d}\bm{\tilde{x}} &= \pi A\left(\int_{k_0}^{k_d} k^{-5/3}\text{d}k + \int_{k_d}^{k_{\text{max}}} k_d^{4/3}k^{-3}\text{d}k\right)\\
&=E_0 A\label{eqn:E0}
\end{align}
We may specify $A$ by requiring the sum of the energy on the coarse grid plus the energy in the eddy initial condition to be conserved in the absence of forcing and dissipation.
In addition, it is natural to include the energy input to the eddies through their interaction with the imposed background shear.
One might also include eddy energy loss to frictional dissipation; however, classical quasigeostrophic turbulence theory suggests that eddy energy loss to friction is negligible so we ignore that effect here.
The above conditions may be satisfied by requiring
\begin{multline}\label{eqn:NRGa}
E_0\frac{1}{2}\frac{\text{d}}{\text{d}t}\iint A\text{d}\bm{x} =k_d^2\iint\left( \ov{\bu_1'\psi_2'}\cdot\nabla\opsi_c-\ov{v_1'\psi_2'}\right)\text{d}\bm{x}\\
-\iint \sum_i\left[-\frac{k_d^2}{2}\nabla\opsi_i\cdot(\ov{\bu_i'\psi_j'}) + \ov{u_i'v_i'}(\pd{x}^2-\pd{y}^2)\opsi_i + \left(\ov{(v_i')^2}-\ov{(u_i')^2}\right)\pd{xy}\opsi_i\right]\text{d}\bm{x}.
\end{multline}
If $A$ were spatially constant but time-varying the above would constrain its evolution.
However, intuition, multiple-scales asymptotics \citep{GSM12}, and simulations \citep{GNS13} suggest that the eddy energy budget can include a great deal of spatial variability.
We therefore adapt the prognostic eddy energy equations proposed by \citet{MA10} and \citet{GSM12} to our situation by requiring $A$ to obey
\begin{multline}
E_0\frac{1}{2}\left(\pd{t}+\bbu_t\cdot\nabla-\nu_A\nabla^2\right) A =k_d^2\left(\ov{\bu_1'\psi_2'}\cdot\nabla\opsi_c - \ov{v_1'\psi_2'}\right)\\
-\sum_i\left[\ov{u_i'v_i'}(\pd{x}^2-\pd{y}^2)\opsi_i - \left(\ov{(v_i')^2}-\ov{(u_i')^2}\right)\pd{xy}\opsi_i\right].\label{eqn:NRG}
\end{multline}
In moving from equation (\ref{eqn:NRGa}) to (\ref{eqn:NRG}), note that $\ov{\bu_1'\psi_2'} = -\ov{\bu_2'\psi_1'}$, which may be derived by the velocity-streamfunction relation $u_i' = -\pd{\tilde{y}}\psi_i'$, $v_i' = \pd{\tilde{x}}\psi_i'$ and integration-by-parts on the eddy domains.
We illustrate the performance of this closure in section \ref{sec:EnergyResults} but much more work remains to explore its properties.

\section{\label{sec:Results}Tests of Stochastic Superparameterization}
\begin{table}
\begin{center}
\caption{\label{table}Time averaged heat flux from all experiments.}
\begin{tabular}{llr}
\toprule
Supercriticality & Closure & Time-Averaged $\iint v_t\psi_c$d$\bm{x}$\\
\midrule
Strong & DNS & $207$\\
Strong & Uncorrelated & $221$ \\
Strong & Deterministic & $176$\\
Moderate & DNS & $23.3$ \\
Moderate & Uncorrelated & $20.1$ \\
Moderate & Correlated, $\eps=12.5$ & $18.9$ \\
Moderate & Correlated, $\eps=25$ & $19.8$\\
Moderate & Correlated, $\eps=25$, $\alpha=1$& $20.9$\\
Moderate & Correlated, $\eps=50$ & $21.4$\\
Moderate & Deterministic & $15.9$\\
Moderate & Prognostic $A$ & $12.6$\\
Weak & DNS & $1.03$\\
Weak & Uncorrelated & $1.09$ \\
Weak & Deterministic & $2.1$ \\
\bottomrule
\end{tabular}
\end{center}
\end{table}

In this section we report the results of low-resolution simulations using the four closures described in sections \ref{sec:ThreeClosures} and \ref{sec:EnergyMethods}, investigating primarily the effects of changing the eddy amplitude $A$ and the cutoff time for eddy evolution $\eps^{-1}$.
Since we have chosen the smallest eddy wavenumber $k_0$ to equal the coarse-grid Nyquist wavenumber, and since the strongest eddy instability (and hence the strongest coupling between large and small scales) is limited to $k<k_d$ we do not consider coarse grids with $100$ or more points in each direction.
(We also note that one can obtain adequate results on grids with $100$ or more points per direction by omitting the eddy terms from the mean equations entirely and tuning the hyperviscosity coefficient $\nu$.)
With the rule-of-thumb that wavenumbers with ten gridpoints per wavelength are well-resolved we also restrict our attention to coarse grids with more than $30$ points in each direction since the most interesting dynamics in our test cases occur for wavenumbers with $k>3$.
Given these constraints we focus on a coarse-grid with $64$ points in each direction; the Nyquist wavenumber for the $64^2$ grid is nearly coincident with the most unstable wavenumber of the linear instability that drives the system (figure \ref{fig:QG_BCI}a), which makes this a particularly difficult test case.

\subsection{The Uncorrelated Stochastic Plane Wave Closure}
\begin{figure}[t!]
\begin{center}
\includegraphics[width=\textwidth]{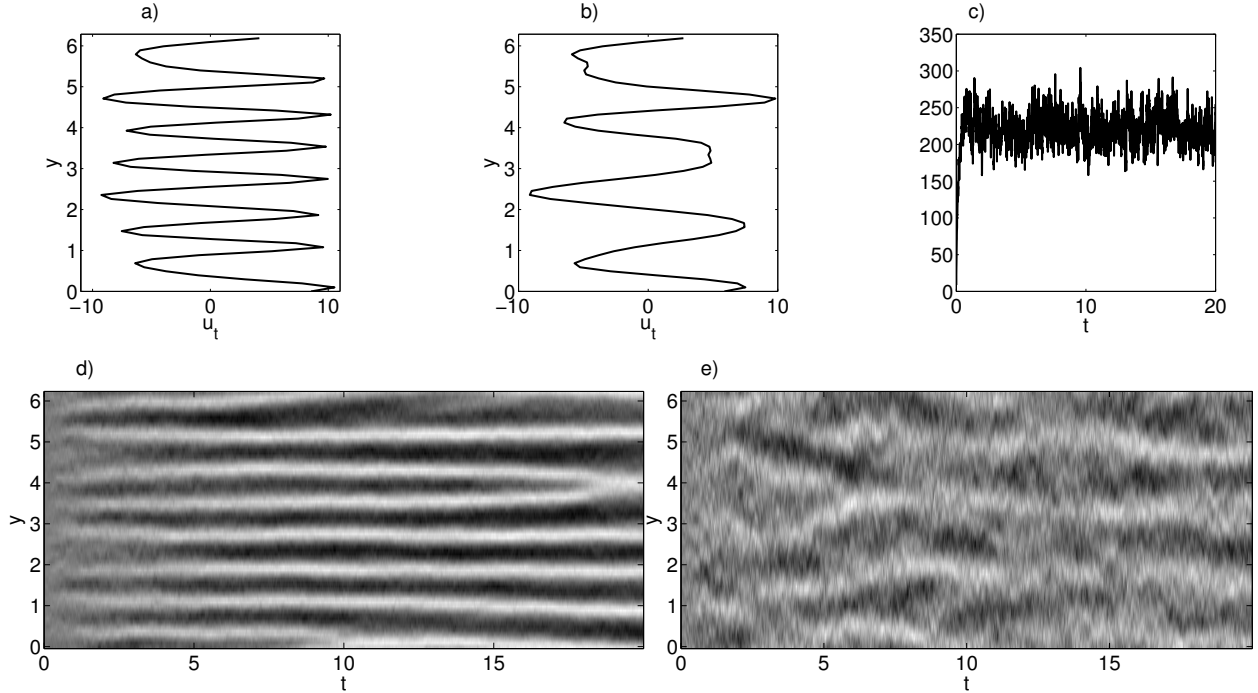}
\caption{Results from the uncorrelated stochastic closure. a) Time- and zonally-averaged zonal barotropic velocity, weakly supercritical case. b) as in a), moderately supercritical case. c) Time series of heat flux in the strongly supercritical case. d) Time series of zonally-averaged zonal barotropic velocity, weakly supercritical case. e) as in d), moderately supercritical case.}\label{fig:Uncorrelated}
\end{center}
\end{figure}

In this section we describe the results of simulations of the mean equations where the eddy terms are approximated using the uncorrelated stochastic plane wave closure described in section \ref{sec:ThreeClosures}, i.e.\,sampling the equilibrium covariance without allowing it to evolve in response to the local mean.
The coarse grid hyperviscous coefficient $\nu$ and eddy amplitude $A$ are tuned by hand to produce optimal results: the weakly supercritical simulation uses $A=1000$ and $\nu=10^{-10}$; the moderately supercritical simulation uses $A=3500$ and $\nu=2\times10^{-10}$; the strongly supercritical simulation uses $A=1.8\times10^4$ and $\nu=4\times10^{-10}$.
These simulations provide a baseline with which to compare the more sophisticated closures.

\begin{figure}
\begin{center}
\includegraphics[width=\textwidth]{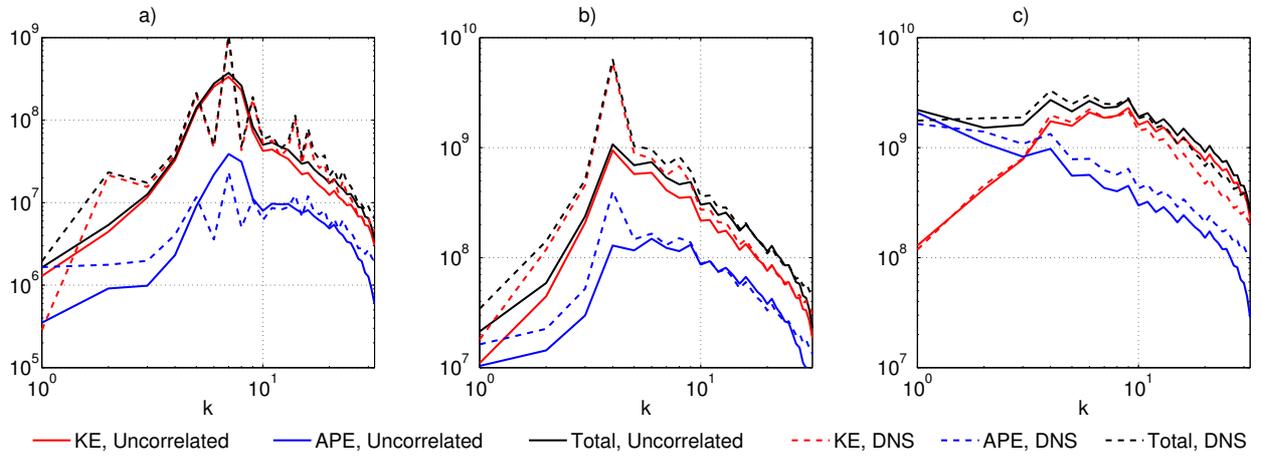}
\caption{Time and angle averaged energy spectra from the uncorrelated stochastic closure.}\label{fig:NullHypSpectra}
\end{center}
\end{figure}

Figures \ref{fig:Uncorrelated}d and \ref{fig:Uncorrelated}e show time series of the zonally-averaged zonal barotropic velocity from the weakly and moderately supercritical simulations, and figures \ref{fig:Uncorrelated}a and \ref{fig:Uncorrelated}b show the time- and zonally-averaged zonal barotropic velocity.
The time series of the meridional heat flux generated by the strongly supercritical simulation is shown in figure \ref{fig:Uncorrelated}c.
Figure \ref{fig:NullHypSpectra} shows the time-averaged energy spectra resulting from the uncorrelated closure; the spectra from the high-resolution reference simulations are shown as dashed lines, while the coarse-resolution spectra are shown as solid lines, with kinetic energy in red, potential in blue, and total in black.

The time-averaged spectra and heat flux for the weakly and strongly supercritical simulations (figure \ref{fig:NullHypSpectra}a and \ref{fig:NullHypSpectra}c, figure \ref{fig:Uncorrelated}c and table \ref{table}) are very similar to the reference simulations; in particular, differences in the precise structure of the spectrum for the weakly supercritical case are within the bounds of variability that result from running the reference simulation with different initial conditions, resulting in jets with slightly different shape (not shown).
The heat flux produced in these simulations matches the high-resolution reference simulations to about $7$\%.
The heat flux in the weakly supercritical experiment in particular is difficult to match because, as shown in figure \ref{fig:Diagnostics}d, the heat flux in the weakly supercritical case is generated primarily by wavenumbers with $k>10$, which are poorly resolved on the coarse grid of $64^2$ points.

Both the high-resolution reference simulation and the uncorrelated closure in the weakly supercritical case have seven jets (figure \ref{fig:QG_BCI}b and \ref{fig:QG_BCI}d, and figure \ref{fig:Uncorrelated}a and \ref{fig:Uncorrelated}d).
This is a striking success since the wavelength of the jets is at the limit of acceptable resolution on the coarse grid.
The jets in the reference simulation are asymmetric with stronger eastward flow with peak amplitude greater than $15$, while the jets from the uncorrelated closure are nearly east/west symmetric with amplitude $10$.
Given the difficulty in representing structures at the limit of acceptable resolution, it is not reasonable to expect the jets in the coarse resolution simulations to exhibit the correct asymmetric east/west structure and amplitude of the high-resolution reference simulation (see figure \ref{fig:DNS}).
We note also that the amplitude of the jets can be increased by tuning $\nu$ and $A$, but only at the expense of changing from seven to six or fewer jets.

In contrast, the four jets in the moderately supercritical simulation are well within the resolving capability of the coarse grid, but are poorly represented by the uncorrelated closure. 
The results shown here are the best obtained from a suite of simulations with $2\times10^{-10}\le\nu\le6\times10^{-10}$ and $3000\le A\le 6500$ (with poor results on the edges of that range), which suggests that better results could not be obtained by further varying the tunable parameters.

\subsection{\label{sec:Correlated}The Correlated Stochastic Plane Wave Closure}
\begin{figure}
\begin{center}
\includegraphics[width=\textwidth]{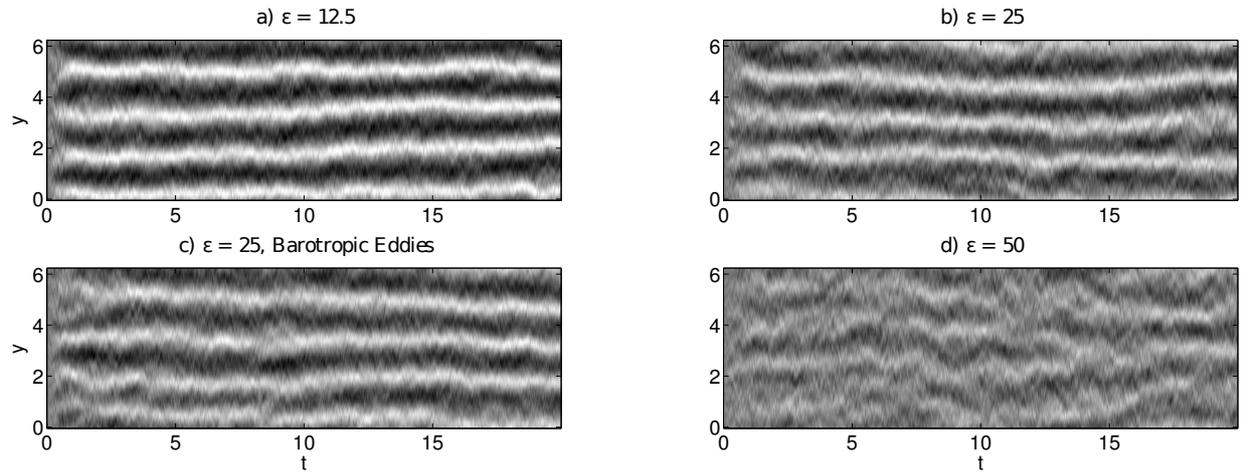}
\caption{Time series of zonally averaged zonal barotropic velocity from the correlated stochastic closure for the moderately supercritical case. a) $\eps=12.5$, b) $\eps=25$, c) $\eps=25$ and $\alpha=1$, d) $\eps=50$. The grayscale is the same in all panels.}\label{fig:Correlated}
\end{center}
\end{figure}

Given the success of the uncorrelated stochastic plane wave closure in the weakly and strongly supercritical simulations, we focus in this section on seeking improvements in the moderately supercritical case through the use of the correlated stochastic plane wave closure.
From a suite of simulations varying $A$, $\nu$, and $\eps$, we find that the optimal values of $A$ and $\nu$ are largely independent of $\eps$; to focus attention on the effect of varying $\eps$ we therefore keep the optimal values of $A=5000$ and $\nu=4\times10^{-10}$ fixed and present results at $\eps=12.5,\,25$, and $50$ using an equilibrium covariance with $\alpha=1/2$ (see section \ref{sec:GC}) and at $\eps=25$ using $\alpha=1$.
\begin{figure}[t!]
\begin{center}
\includegraphics[width=\textwidth]{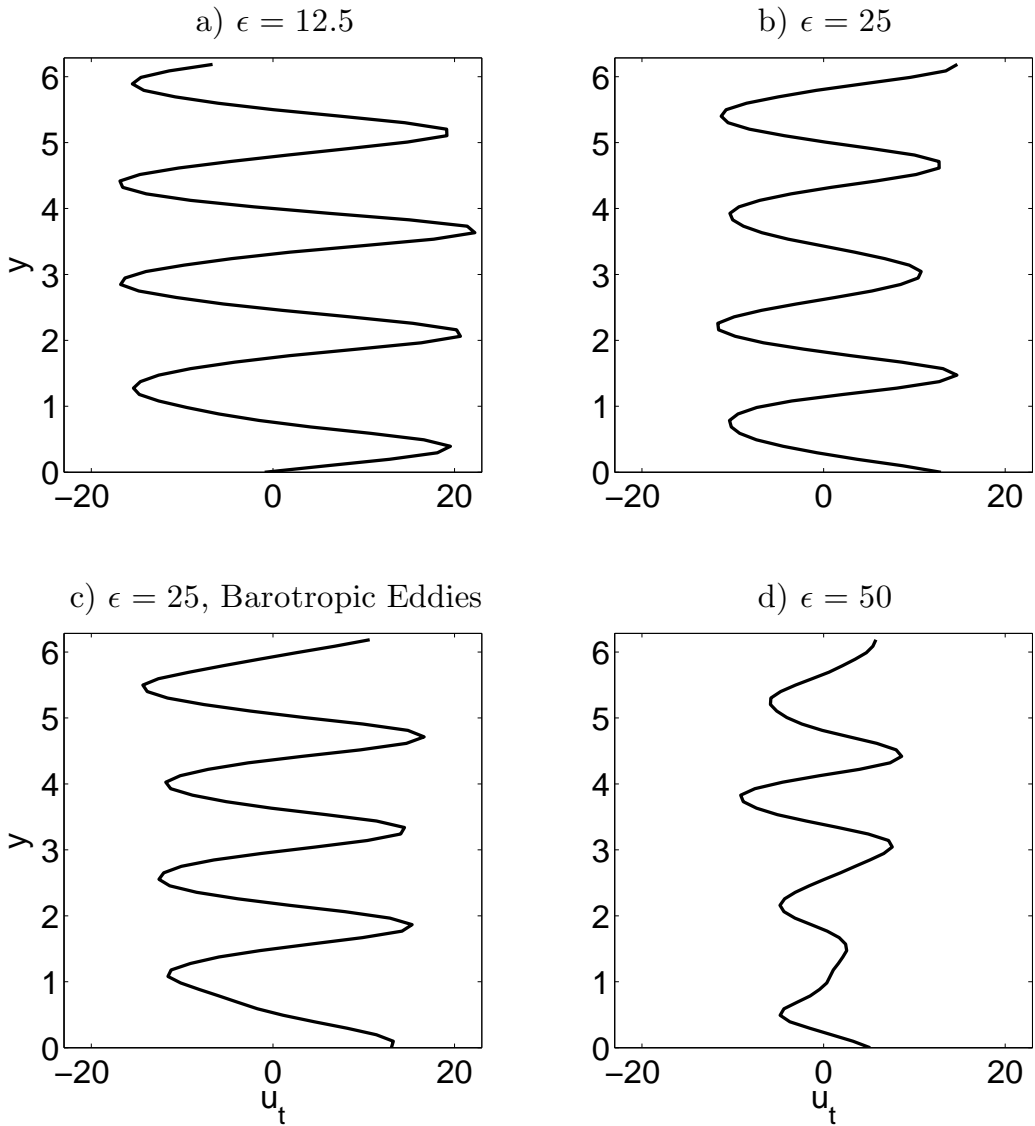} 
\caption{Time and zonally averaged zonal barotropic velocity from the correlated stochastic closure in the moderately supercritical case; a)-d) as in figure \ref{fig:Correlated}.}\label{fig:CorrLine}
\end{center}
\end{figure}

\begin{figure}[t!]
\begin{center}
\includegraphics[width=\textwidth]{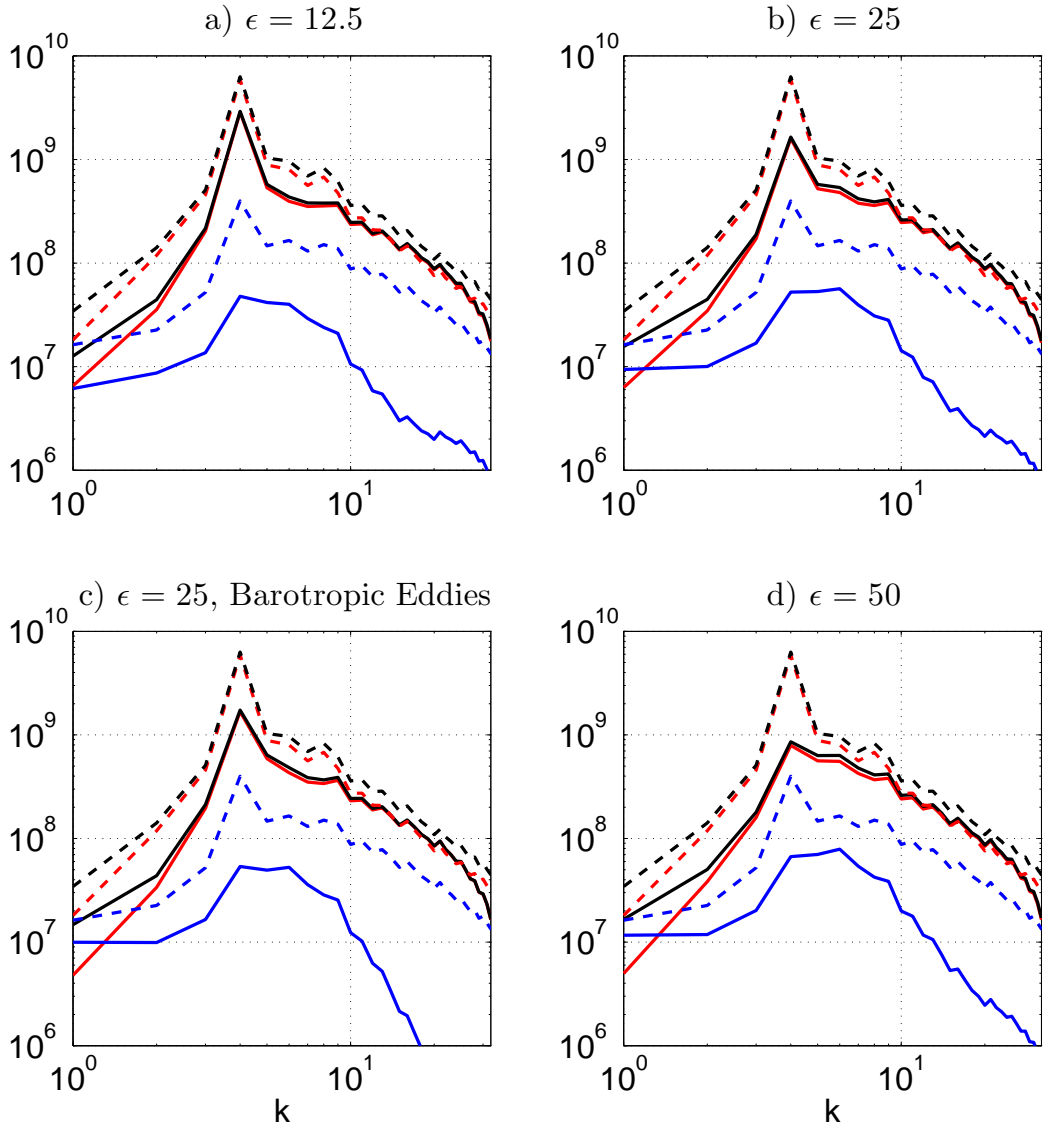}
\caption{Time and angle averaged energy spectra from the correlated stochastic closure in the moderately supercritical case; a)-d) as in figure \ref{fig:Correlated} and legend as in figure \ref{fig:NullHypSpectra}.}\label{fig:CorrSpectra}
\end{center}
\end{figure}

Figure \ref{fig:Correlated} shows time series of the zonally-averaged zonal barotropic velocity from the four simulations varying $\eps$ and $\alpha$; plots of the time-averaged jet structure are shown in figure \ref{fig:CorrLine}.
Table \ref{table} presents the heat flux for each experiment, and figure \ref{fig:CorrSpectra} shows the time-averaged energy spectra.
We begin by noting that the jet structure is clearly improved at smaller $\eps$, with little difference between $\alpha=1/2$ and $\alpha=1$.
The jet structure at $\eps=12.5$ is better than at $\eps=25$ or $50$, and far better than using the uncorrelated closure, with amplitude of approximately $20$ and some indication of east/west asymmetry. 
The clear improvement in the jet structure as $\eps$ decreases suggests that further improvements might be had by further decreasing $\eps$, but there is a tradeoff between improved jet structure and worsened heat flux as $\eps$ decreases: as shown in table \ref{table}, the heat flux is optimal at $\eps=50$ and decreases with $\eps$.
Nevertheless, the heat flux is correct to within $20$\% in all cases, whereas the jet amplitude is too low by about $33$\%.
Figure \ref{fig:CorrSpectra}, which shows the time-averaged energy spectra from the four experiments, also indicates that the potential energy spectrum, which is too low in all simulations, is best at $\eps=50$. 
We note though, that there is much less potential energy than kinetic energy, so the degradation of the potential energy spectrum is less important than the improvement in the peak of the kinetic energy spectrum.
Note also that the potential energy spectrum is slightly better at small scales with $\alpha=1/2$ than with $\alpha=1$.
This is because the Reynolds stress terms with $\alpha=1/2$ are more baroclinic than with $\alpha=1$, and therefore backscatter more energy into the potential energy spectrum, slightly decreasing the efficiency of the already too-efficient downscale cascade of potential energy.

\subsection{\label{sec:Deterministic}The Deterministic Closure}
In this section we report results of the deterministic closure described in section \ref{sec:ThreeClosures}.
We emphasize that the deterministic closure should not be expected to work well in this situation with an inverse cascade mediated by stochastic backscatter.
Since the closure has a significantly different character than the stochastic closures we present results from all three parameter regimes.
We begin by noting that the deterministic closure does not require the relatively large values of $\nu$ needed to stabilize the stochastic closures; all experiments reported in this section use $\nu=10^{-12}$, and the results are not very sensitive to variations of $\nu$.
Also, because the deterministic closure is decoupled from the length of the coarse grid time step, we use the same adaptive time step algorithm as in the reference simulations rather than a fixed time step as used for the stochastic closures.
While we have run tests with $\eps=12.5$, $25$, $50$ and $100$, the results obtained at $\eps=25$ are representative of the general behavior, and only the latter are presented at weak and moderate supercriticalities.
\begin{figure}
\begin{center}
\includegraphics[width=\textwidth]{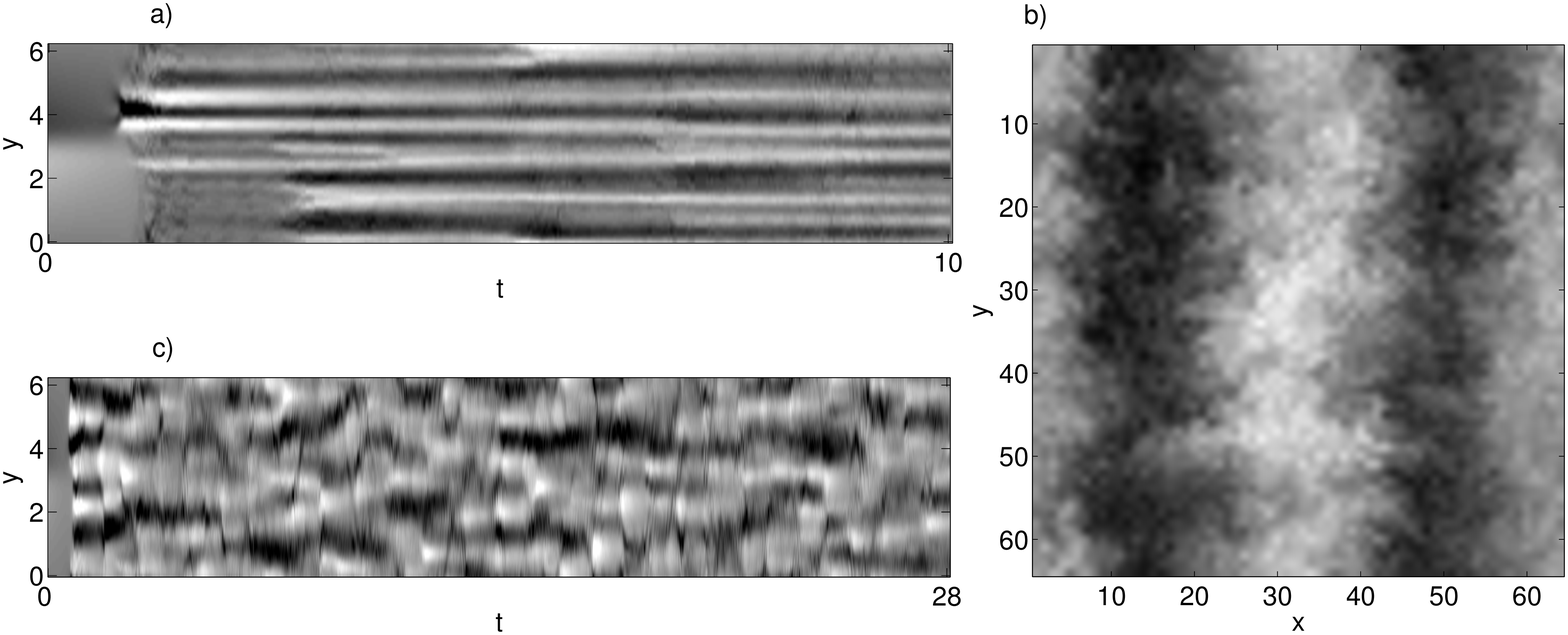}
\caption{Results from the deterministic closure. a) Time series of zonally averaged zonal barotropic velocity, weakly supercritical. b) Snapshot of upper layer potential vorticity $\oq_1$, strongly supercritical. c) as in a), moderately supercritical.}\label{fig:DetImage}
\end{center}
\end{figure}

In the strongly supercritical case the simulations at $\eps\le25$ are numerically unstable for all values of $A$ and $\nu$ tested.
We propose the following explanation.
At strong supercriticality the range of baroclinic mean shear observed on the coarse grid is much broader than at weak and moderate supercriticality.
At small $\eps$ the eddies have a long time to respond to the strongly unstable mean shear, even when those values of mean shear exist only briefly on the coarse grid, and the eddies produce an unrealistically large response.
These large eddy terms then require an extremely small time step to accommodate them, and the use of moderate time steps results in instability.
The numerical conditioning of stochastic SP was improved in the toy model of \citet{MG09} by truncating the growth of overly-unstable eddy modes during the quadrature used to compute the eddy terms.
As an alternative, the conditioning could be improved by truncating the values of the mean shear used to generate the eddy terms, so that anomalously large values of local mean shear are reduced before being used to calculate the eddy response.
Again, the conditioning could be improved by increasing the damping rate $\gamma_0$ so that the instabilities do not develop as quickly.
Since the deterministic closure is included here primarily for illustration we do not pursue these options further, but simply present the best stable results for the strongly supercritical case at $\eps=50$.

The interpolant used to compute the eddy terms at moderate supercriticality is the same as used in the simulations with the correlated stochastic closure.
For the weakly supercritical case the interpolant is computed in the range $|\bm{\hat{k}}\cdot\bbU_c|\le 3.5$, $|\bm{\hat{k}}\times\nabla\ov{\omega}_c|\le 10^3$, and $|\bm{\hat{k}}\times\nabla\ov{\omega}_t|\le 7\times10^3$ with $\alpha=0.25$, and for the strongly supercritical case the interpolant is computed in the range $|\bm{\hat{k}}\cdot\bbU_c|\le 7.5$, $|\bm{\hat{k}}\times\nabla\ov{\omega}_c|\le 10^4$, and $|\bm{\hat{k}}\times\nabla\ov{\omega}_t|\le 5\times10^4$ with $\alpha=0.5$.
We note that the range of baroclinic shear used to compute the interpolant at strong supercriticality is slightly less than the range of observed values; as suggested above, this truncation improves the numerical conditioning.

For each scenario, the value of $A$ is hand tuned to produce optimal results; the weakly supercritical case uses $A=10^4$, the moderately supercritical case uses $A=2\times10^4$, and the strongly supercritical case uses $A=10^3$.
Figure \ref{fig:DetImage}a and \ref{fig:DetImage}c show time series of the zonally-averaged zonal barotropic velocity in the weakly and strongly supercritical experiments, respectively.
The jets in both cases are far too weak, with amplitude approximately $4$ and $8$, respectively, and both exhibit unrealistic temporal variation. 
Figure \ref{fig:DetImage}b shows a snapshot of the upper layer potential vorticity $\ov{q}_1$ from the strongly supercritical simulation.
A strong mode-one Rossby wave is clearly evident; this strong mode-one behavior is typical of the simulations at strong supercriticality and is entirely spurious since it is not evident in the reference simulations.
\begin{figure}
\begin{center}
\includegraphics[width=\textwidth]{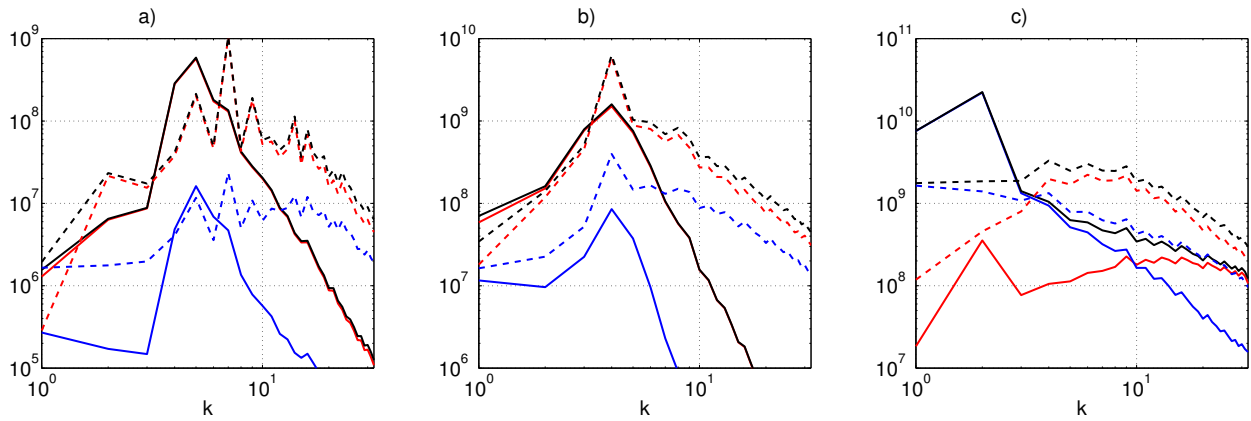}
\caption{Time and angle averaged energy spectra from the deterministic closure. Legend as in figure \ref{fig:NullHypSpectra}.}\label{fig:DetSpec}
\end{center}
\end{figure}

The time-averaged energy spectra from the deterministic closure experiments are shown in figure \ref{fig:DetSpec}.
Although the spectra are largely incorrect in all cases, it is worth noting that, unlike negative viscosity, the closure is able to inject energy to the large scales and remain computationally stable despite being completely deterministic.
Second, the extremely steep kinetic energy spectra in the weakly and moderately supercritical experiments suggest that the closure is also smoothing the small scales, since the steepness of the spectrum is clearly not due to the weak effect of hyperviscosity.
Finally, table \ref{table} shows that the heat flux generated by the deterministic closure is significantly incorrect.

\subsection{\label{sec:EnergyResults}Closure with Prognostic Eddy Energy Density}
In this section we briefly report on results obtained using the closure described in section \ref{sec:EnergyMethods}.
This closure is similar to the correlated stochastic closure, except that the eddy amplitude $A$ obeys the heuristically-motivated prognostic partial differential equation (\ref{eqn:NRG}) instead of being a tunable constant.
We again focus on the moderately supercritical scenario since the uncorrelated stochastic closure leaves little room for improvement in the other two scenarios.
Recall that in the equation for $A$, equation (\ref{eqn:NRG}), the net eddy energy equilibrates by balancing energy input from interaction with the imposed background shear and from absorption of the downscale potential energy cascade with energy output to the inverse kinetic energy cascade on the coarse grid.

Results are shown here using $\eps=25$ and $\gamma_0=1$ where $\gamma_0$ is the damping coefficient in (\ref{eqn:gamma0}).
The results at $\eps=12.5$ and $\gamma_0=30$ are similar, but have lower overall energy and weaker jets.
We hypothesize that this results from insufficient energy injection to the eddies from the imposed background, since $\gamma_0=30$ is sufficient to stabilize the linear instability of the imposed background shear.
To counter this effect we used $\eps=12.5$ and $\gamma_0=15$, but the algorithm was numerically unstable and developed negative values of $A$.
As a compromise we ended with $\eps=25$ and $\gamma_0=1$, which increases the overall energy compared to $\gamma_0=30$.
We also set $\nu=6\times10^{-10}$, which improves the results slightly.
The parameters $\nu_A$ and $E_0$ used in specifying the evolution equation (\ref{eqn:NRG}) have the values $\nu_A=1$ with $E_0$ defined in (\ref{eqn:E0}). 
Results are shown for $\nu_A=1$; for much smaller values of $\nu_A<0.1$ the simulations can develop negative $A$, which is clearly unrealistic, and become unstable.

\begin{figure}[t!]
\begin{center}
\includegraphics[width=\textwidth]{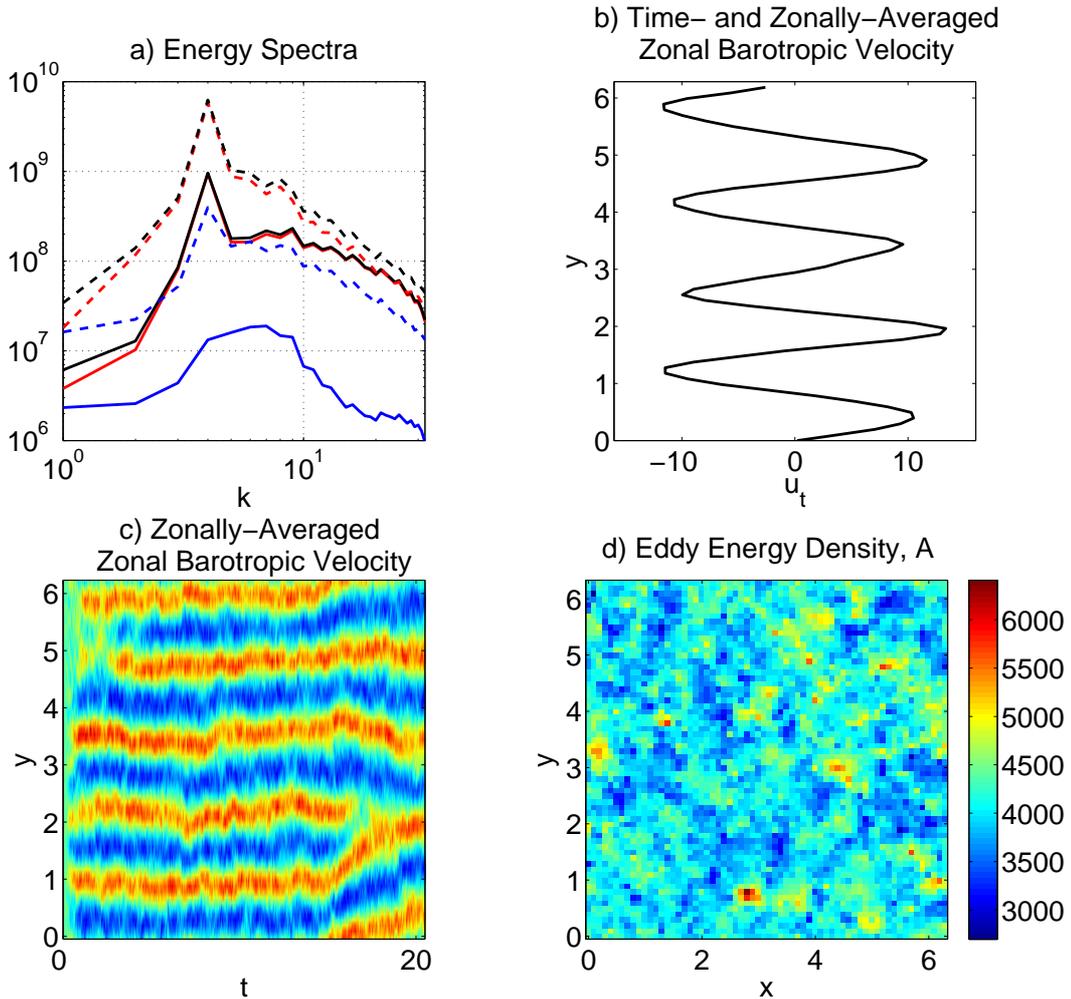}
\caption{Results from the prognostic-$A$ closure. a) Time and angle averaged energy spectra; legend as in figure \ref{fig:NullHypSpectra}. b) Time and zonally averaged zonal barotropic velocity. c) Time series of zonally averaged zonal barotropic velocity. d) Snapshot of $A$.}\label{fig:Energy}
\end{center}
\end{figure}

Figure \ref{fig:Energy} shows the time-averaged energy spectra (\ref{fig:Energy}a), the time- and zonally-averaged zonal barotropic velocity (\ref{fig:Energy}b), a time series of the zonally-averaged zonal barotropic velocity (\ref{fig:Energy}c), and a snapshot of $A$ (\ref{fig:Energy}d).
The spatial-average value of $A$ equilibrates to approximately $4000$.
Recalling that the optimal results for the correlated stochastic closure were obtained for $A=5000$, the prognostic closure is generating too little eddy energy.
As a result, the jets are too weak, and the meridional flow on the large scales is too weak to generate sufficient heat flux (see table \ref{table}).
The potential energy spectrum is also much too low, similar to the results from the correlated stochastic closure with constant $A$.
Although the results are not perfect, we are hopeful that further development of this closure will enable $A$ to be computed as part of the solution instead of being an externally tuned parameter.
\section{Discussion\label{sec:Discussion}}
In this section we discuss the stochastic SP algorithms the results of the foregoing experiments, and compare briefly with other stochastic parameterization methods.
More information on non-stochastic SP can be found in \citet{MG13}.

Stochastic SP begins by formally separating the master equations that govern the dynamics at all scales into mean and eddy equations by means of a Reynolds average.
The point approximation is then applied, which re-interprets the eddy equation as evolving in subdomains embedded at each point of the physical domain, and re-interprets the Reynolds average as an average over the embedded domains.
This imposes a scale separation between the mean and the eddies and produces a set of multiscale equations amenable to solution by SP.
The point approximation is reminiscent of multiple-scales asymptotics (e.g.\,\citep{GSM12}) in that it produces eddy equations that evolve in pseudo-physical domains embedded in the physical domain, but the point approximation differs from multiple-scales asymptotic methods primarily in that it does not assume extreme/asymptotic scale separation.

Stochastic SP then replaces the nonlinear eddy equations that result from the application of the point approximation by quasi-linear stochastic partial differential equations.
The eddies are modeled as spatially homogeneous Gaussian random functions in the embedded domains (although they are not homogeneous across the physical domain).
The properties of the Gaussian closure of the eddies are set by basic phenomenological considerations.
Although these properties are compared to the results of high-resolution reference simulations, they are not tuned to match those simulations, nor are they especially good approximations of those simulations, as evidenced by the fact that the curves in figure \ref{fig:Diagnostics}a-c are not flat.
Despite the inexactness of the match between the details adopted for the Gaussian closure and the properties of the reference simulations, it is clear that some \emph{a priori} knowledge of the structure of the eddy spectra and cross-spectra is required.
In cases where such knowledge is lacking, the simple Gaussian closure model adopted here could be replaced by more sophisticated models, based for example on the modified quasilinear Gaussian approach of \citet{SM13}, or the more complicated EDQNM (eddy damped quasi-normal Markovian) turbulence models \citep{Lesieur08}; while this would significantly increase the cost and complexity it would diminish the required amount of \emph{a priori} information about the small scales, and could potentially improve the accuracy of the eddy model.

The feedback from the eddies to the mean appears as the average of products of eddy variables over the embedded domains.
Since the eddies are homogeneous on the embedded domains, the spatial average is equivalent to an ensemble average and the feedback terms are not stochastic.
Such a closure was used in \cite{GM14}; in the present approach, following GM, we alter the eddy model by introducing random anisotropy to the eddy model: the eddies are made to consist entirely of a sum of Fourier modes in a single direction that is a random function over the large-scale physical domain.
This element of random anisotropy is a minimalistic model corresponding to the simple fact that in reality the small scale eddies in a given location are typically not isotropic, but consist of a few randomly oriented vortices, waves, and filaments.
The details of the random eddy orientation algorithm adopted here leave room for improvement, for example by incorporating realistic spatiotemporal correlations into the local eddy direction.
This is a potential topic of future research.

The foregoing tests show that the stochastic SP algorithms developed here can reproduce the energy spectra, heat flux, and jet structure of the reference simulations reasonably well.
The main difficulties are in producing accurate potential energy spectra since the algorithms drain too much large-scale potential energy.
A particular concern is the fact that the eddy amplitude $A$ is tuned to reproduce the results of the reference simulations; section \ref{sec:EnergyMethods} provides a model for $A$ based on energy conservation, but the results of section \ref{sec:EnergyResults} show that further work is needed.

The results are also sensitive to the value of the coarse-grid viscosity $\nu$, though sensitivity to $\nu$ is less than to $A$: changing the viscosity by a factor of two from $2\times10^{-10}$ to $4\times10^{-10}$ has less impact than changing $A$ by twenty percent from $5000$ to $4000$.
This sensitivity is related to the fact that the eddy angle is uncorrelated in space and time, which leads to extremely rough eddy fluxes.
As mentioned above, adding spatial and temporal correlations to the eddy angle is a clear avenue of possible improvement for future research, and adding such correlations will likely decrease the sensitivity to the coarse-grid viscosity $\nu$.
The results are sensitive to the time over which the eddies are allowed to respond to the local mean conditions, $\eps^{-1}$.
Again, sensitivity to $\eps$ is lower than sensitivity to $A$: changing $\eps$ by a factor of two from $12.5$ to $25$ has less impact than changing $A$ by twenty percent (see figures \ref{fig:Correlated}, \ref{fig:CorrLine}, and \ref{fig:CorrSpectra}).
As noted above, if some information on the decorrelation time of the large scales is available, then it would seem reasonable to choose $\eps$ such that $\eps^{-1}$ is comparable to or shorter than the decorrelation time of the large scales.

\section{Conclusions\label{sec:Conclusions}}
In this article we expand and develop the work of \citet{GM13a} in applying stochastic superparameterization to quasigeostrophic turbulence.
Stochastic parameterization techniques can be broadly classified into two categories: those that rely on stochastic models of the unresolved eddy dynamics, and those that develop stochastic models only of the effects of the  unresolved eddies on the resolved large scales.
The current work and the work of \citet{FOZ12} (and citations therein) and \citet{KBM10} are examples of the former category, while the work of \citet{L90}, \citet{BSLP09}, and \citet{B13} are examples of the latter.
The stochastic parameterization algorithms developed here are successful in generating robust inverse cascades of kinetic energy and simultaneous downscale cascades of potential energy on coarse computational grids that do not resolve the deformation scale.
The test scenario considered here is particularly difficult since the unresolved small scales act as the primary source of kinetic energy for the large scales, unlike typical scenarios where the largest scales are directly forced by buoyancy and momentum fluxes.
The use of random-direction reduced-dimensional embedded domains, together with the stochastic superparameterization framework (point approximation and Gaussian closure) results in extremely efficient seamless stochastic SP algorithms that do not require simulation of nonlinear eddy PDEs on small scale grids as in conventional superparameterization.

For stochastic superparameterization in general, our results show that the multiscale formulation can still be effective in situations without scale separation, although the algorithm is naturally expected to work even better in situations with scale separation.
We find that when the eddies have no memory of their previous state, but are instead re-set to an equilibrium initial condition at the beginning of each large-scale time step (more precisely, the eddies are re-set to an initial condition each time the eddy terms are evaluated) the evolution timescale for the eddies $\eps^{-1}$ should be decoupled from the size of the coarse-grid time step to allow the eddies sufficient time to respond to the properties of the local mean.
An experimental closure based on energy conservation ideas is developed to show how some memory may be retained by the eddies even when they are continually being re-set to an initial condition.

Future directions include coupling the QG eddy model to a coarse-grid ocean general circulation model (GCM), which will require a more sophisticated treatment of the vertical direction and some care in setting up the eddy equations using the point approximation.
Other issues not addressed here include the effects of bottom topography and lateral boundaries; the latter having never been addressed in a superparameterization context, to our knowledge.
The framework of stochastic superparameterization used here may also be applicable in other geophysical settings, including thermal convection in the ocean mixed layer.
A general formulation of the framework of stochastic superparameterization can be found in \citet{MG13}.

Another direction for future research is the development of statistical prediction and state estimation algorithms using superparameterization.
Stochastic parameterization in general and stochastic SP in particular are well suited to improving the efficiency of ensemble-based prediction and state-estimation algorithms \cite{HM13}.
In geophysical applications the state of the unresolved scales is typically unconstrained by observation, and prediction systems attempt to solve for the most probable evolution of the large scales.
High resolution simulations are often too costly to use with a sufficiently large ensemble size, but low resolution simulations often display insufficient variability, which can lead to filter divergence (i.e.\,observations are ignored).
Stochastic SP offers a strategy for developing efficient low resolution simulations with increased variability due to the stochastic eddy forcing.
These properties suggest that it will improve the performance of ensemble-based prediction systems.

\section*{Acknowledgements}
I.G.~is supported as a postdoctoral fellow by the NSF \emph{Collaborations in Mathematical Geosciences} program, grant DMS-1025468.
A.J.M.~is partially supported by the Office of Naval Research grants N00014-11-1-0306 and ONR-DRI N0014-10-1-0554.

\appendix
\section{Miscellaneous Formulas \label{sec:Appendix}}
The formulas for the remaining eddy terms as integrals over $\bm{k}$ are
\begin{align*}
\ov{u_2'(\psi_1'-\psi_2')} &= -\ov{u_1'(\psi_2'-\psi_1')}\\
\notag
\ov{v_1'(\psi_2'-\psi_1')}=\ov{v_1'\psi_2'}&=-i\iint\left(\eps\int_0^{\eps^{-1}}\mathbb{E}\left[k_x\hat{\psi}_1^*\hat{\psi}_2\right]\text{d}\tau\right)\text{d}\bm{k}\\
&=-\iint k_x\left(\eps\int_0^{\eps^{-1}}\mathbb{E}\left[\mathcal{I}\{\hat{\psi}_1\hat{\psi}_2^*\}\right]\text{d}\tau\right)\text{d}\bm{k}\\
\ov{v_2'(\psi_1'-\psi_2')} &= -\ov{v_1'(\psi_2'-\psi_1')}\\
\ov{(v_i')^2}-\ov{(u_i')^2} &= \iint (k_x^2-k_y^2) \left(\eps\int_0^{\eps^{-1}}\mathbb{E}\left[|\hat{\psi}_{i}|^2\right]\text{d}\tau\right)\text{d}\bm{k}.
\end{align*}
In polar form these are
\begin{align}
\ov{u_2'(\psi_1'-\psi_2')} &= -\ov{u_1'(\psi_2'-\psi_1')}\label{eqn:ub2Polar}\\
\ov{v_1'(\psi_2'-\psi_1')}&=-\int_0^{2\pi}\int_{k_0}^{k_{\text{max}}} k^2\cos(\theta)\left(\eps\int_0^{\eps^{-1}}\mathbb{E}\left[\mathcal{I}\{\hat{\psi}_1\hat{\psi}_2^*\}\right]\text{d}\tau\right)\text{d}k\text{d}\theta\label{eqn:vbPolar}\\
\ov{v_2'(\psi_1'-\psi_2')} &= -\ov{v_1'(\psi_2'-\psi_1')}\\
\ov{(v_i')^2}-\ov{(u_i')^2} &= \int_0^{2\pi}\int_{k_0}^{k_{\text{max}}} k^3\cos(2\theta) \left(\eps\int_0^{\eps^{-1}}\mathbb{E}\left[|\hat{\psi}_i|^2\right]\text{d}\tau\right)\text{d}k\text{d}\theta\label{eqn:vmuPolar}.
\end{align}

The linear propagator in equation \ref{eqn:bM} has the form
\begin{equation}
\text{\bf M}_k = \frac{1}{4 k^2(k^2+k_d^2)}\tilde{\text{\bf M}}_{\bm{k}}-\frac{r}{2(k^2+k_d^2)}\text{\bf R}_{\bm{k}}- 2(\gamma_k+\nu k^8)\text{\bf I}
\end{equation}
where {\bf I} is the identity matrix,
\begin{equation}
\tilde{\text{\bf R}}_{\bm{k}} =\left[ \begin{array}{c c c c}
0 &2 k_d^2 & 0 & 0 \\
0 & 2 k^2 + k_d^2 & 0 &k_d^2 \\
0 & 0 &  2 k^2 + k_d^2 & 0 \\
0 & 0 & 0 &  2(2 k^2 + k_d^2) 
\end{array}\right]
\end{equation}
 and $\tilde{\text{\bf M}}_{\bm{k}}$ has the following nonzero entries
\begin{align}
\left\{\tilde{\text{\bf M}}_{\bm{k}}\right\}_{3,1} &= k_d^2 (-2 \bm{k}\times\nabla\ov{Q}_1 + 2(2 k^2 + k_d^2) \bbU_c\cdot\bm{k})\\
\left\{\tilde{\text{\bf M}}_{\bm{k}}\right\}_{1,3} &=2 k_d^2 (2 \bm{k}\times\nabla\ov{Q}_2 + 2(2 k^2 + k_d^2) \bbU_c\cdot\bm{k})\\
\left\{\tilde{\text{\bf M}}_{\bm{k}}\right\}_{3,2} &=4 (2 k^2 + k_d^2)\bm{k}\times\nabla \ov{Q}_c - 4 (2 k^4 + 2 k^2 k_d^2 + k_d^4) \bbU_c\cdot\bm{k}\\
\left\{\tilde{\text{\bf M}}_{\bm{k}}\right\}_{2,3} &=-4 (2 k^2 + k_d^2)\bm{k}\times\nabla \ov{Q}_c + 4 (2 k^4 + 2 k^2 k_d^2 + k_d^4) \bbU_c\cdot\bm{k}\\
\left\{\tilde{\text{\bf M}}_{\bm{k}}\right\}_{4,3} &=2 k_d^2 (-2 \bm{k}\times\nabla\ov{Q}_1 +2 (2 k^2 + k_d^2) \bbU_c\cdot\bm{k})\\
\left\{\tilde{\text{\bf M}}_{\bm{k}}\right\}_{3,4} &=k_d^2 (2 \bm{k}\times\nabla\ov{Q}_2 +2 (2 k^2 + k_d^2) \bbU_c\cdot\bm{k})
\end{align}
The linear propagator $\bM_{\bm{k}}$ depends only on the baroclinic part of the velocity $\bbU_c = (\bbu_1-\bbu_2)/2 + \bm{\hat{x}}$; this is not surprising since the barotropic velocity amounts to a uniform translation and does not affect the eddy covariance.

\bibliographystyle{model1-num-names}
\bibliography{GM_JCP_Draft_v3}
\end{document}